\def\tr{{\rm tr}\,}
\def\Tr{{\rm Tr}\,}
\def\wt{\widetilde}
\def\sgn{{\rm sgn\,}}
\def\b{\bibitem}
\def\be{\begin{equation}}
\def\ee{\end{equation}}
\def\bea{\begin{eqnarray}}
\def\eea{\end{eqnarray}}
\def\bml{\begin{mathletters}}
\def\eml{\end{mathletters}}
\documentstyle[aps,prb,eqsecnum,psfig,epsf,floats]{revtex}
\draft
\begin{document}
\def\SNG{{\em Physical Review Style and Notation Guide}}
\def\LUG {{\em \LaTeX{} User's Guide \& Reference Manual}}
\def\btt#1{{\tt$\backslash$\string#1}}%
\def\REVTeX{REV\TeX}
\def\AmS{{\protect\the\textfont2
        A\kern-.1667em\lower.5ex\hbox{M}\kern-.125emS}}
\def\AmSLaTeX{\AmS-\LaTeX}
\def\BibTeX{\rm B{\sc ib}\TeX}
\twocolumn[\hsize\textwidth\columnwidth\hsize\csname@twocolumnfalse%
\endcsname
\title{Local field theory for disordered itinerant quantum ferromagnets\\
       \small{$[$ Phys. Rev. B {\bf 63}, 174427 (2001) $]$}
                }
\author{D. Belitz}
\address{Department of Physics and Materials Science Institute\\
         University of Oregon,
         Eugene, OR 97403}
\author{T.R. Kirkpatrick}
\address{Institute for Physical Science and Technology, and Department of 
         Physics\\
         University of Maryland,
         College Park, MD 20742}
\author{Maria Teresa Mercaldo}
\address{Institute for Physical Science and Technology, and Department of
         Physics\\
         University of Maryland,
         College Park, MD 20742\\
         and\\
         Dipartimento di Scienze Fisiche ``E.R. Caianiello'' and Istituto
         Nazionale di Fisica per la Materia,\\
         Universit{\'a} di Salerno, I-84081 Baronissi (SA), Italy}
\author{Sharon L. Sessions}
\address{Department of Physics and Materials Science Institute\\
         University of Oregon,
         Eugene, OR 97403}
\date{\today}
\maketitle

\begin{abstract}
An effective field theory is derived that describes the quantum critical
behavior of itinerant ferromagnets in the presence of quenched disorder. 
In contrast to previous
approaches, all soft modes are kept explicitly. The resulting effective
theory is local and allows for an explicit perturbative treatment.
It is shown that previous suggestions for the critical fixed point and
the critical behavior are recovered under certain assumptions. The
validity of these assumptions is discussed in the light of the existence
of two different time scales. It is shown that, in contrast to previous
suggestions, the correct
fixed point action is not Gaussian, and that the previously
proposed critical behavior was correct only up to logarithmic corrections.
The connection with other theories of disordered interacting electrons, 
and in particular with the resolution of the
runaway flow problem encountered in these
theories, is also discussed.
\end{abstract}
]
\section{Introduction}
\label{sec:I}

The theory of quantum phase transitions, i.e. phase transitions at zero
temperature $(T=0)$, is an important problem that is relevant to
many topics in condensed matter physics.\cite{Sachdev_book}
Perhaps the most obvious example is the $T=0$ transition from a paramagnet
to an itinerant ferromagnet as it occurs in, e.g., diluted Ni, or in
solid solutions like MnSi. Historically, this was the first quantum phase
transition that was studied in detail. Hertz\cite{Hertz} 
showed how to treat this transition by means of renormalization group
(RG) methods, and he concluded that the transition in the physically
interesting dimension $d=3$ was mean-field like. This conclusion hinged
on the observation that in quantum statistical mechanics the effective
dimension of a system for scaling purposes is $d+z$, with $d$ the spatial
dimensionality and $z$ the dynamical critical exponent. Since in a simple
theory at tree level one has $z=3$ for clean itinerant quantum ferromagnets, 
and $z=4$ for disordered ones,\cite{Hertz} 
this seemed to imply that the
upper critical dimension $d_c^+$, above which one finds mean-field critical
behavior, is $d_c^+ = 1$ and $d_c^+ = 0$ in the clean and disordered cases,
respectively. 

This conclusion was later
challenged.\cite{Sachdev,us_dirty,us_clean} Reference \onlinecite{Sachdev}
noted that Hertz's results for clean systems in $d=1-\epsilon$ dimensions 
were inconsistent with general scaling arguments. This left open the 
possibility of mean-field behavior in physical dimensions. However, in 
Refs.\ \onlinecite{us_dirty,us_clean}
it was shown that the critical behavior in $d>1$ (clean case) and $d>0$
(disordered case), respectively, is not mean-field like after all. The
salient point is that in itinerant electron systems at $T=0$,
soft modes other than the order parameter fluctuations exist. These modes are
diffusive (in disordered systems) or ballistic (in clean ones) particle-hole
excitations. Since they couple to the order parameter
fluctuations, they influence the critical behavior. Specifically, they
lead to an effective long-range interaction between the order parameter
fluctuations. As a result, Refs.\ \onlinecite{us_dirty,us_clean} found
that the critical behavior is governed by a Gaussian fixed point which,
however, does not yield mean-field exponents. The critical behavior determined
in these references was claimed to be exact.

A separate, and seemingly unconnected, development in the many-electron
problem has been the study of metal-insulator transitions of disordered
interacting electrons.\cite{us_R} For one of the universality classes that
occur in this problem, a transition was found that is {\em not} a
metal-insulator transition, but rather of magnetic 
nature.\cite{runaway_footnote} While the order parameter, and the nature of 
the ordered phase, could not be identified with the methods employed,
the critical behavior for all quantities other than the order parameter
was determined.\cite{us_IFS} Apart from differences in logarithmic 
corrections to
power laws, this critical behavior turned out to be identical with the
Gaussian critical behavior for the disordered ferromagnetic transition.
This led, in Ref.\ \onlinecite{us_dirty}, to the suggestion that the
unidentified transition studied in Ref.\ \onlinecite{us_IFS} was the
ferromagnetic transition. The discrepancy with respect to the logarithmic
terms was explained as due to the fact that of the two integral equations
in Ref.\ \onlinecite{us_IFS} only one had been shown to be exact. The proposal
thus was that the two approaches describe the same transition, and that the
critical behavior found in Ref.\ \onlinecite{us_dirty} was exact while the
one in Ref.\ \onlinecite{us_IFS} represented an approximation.

The theory developed in Refs.\ \onlinecite{us_dirty,us_clean} suffers from one
major drawback: Since the additional soft modes were integrated out in order
to obtain a description entirely in terms of the order parameter, the effective
field theory that was derived is nonlocal\cite{local_footnote}
and not very suitable for
perturbatively calculating effects that depend on all of the soft modes in the
system. The analysis in 
Refs.\ \onlinecite{us_dirty,us_clean} therefore was restricted to
power counting arguments
at tree level to show that all non-Gaussian terms are irrelevant in a 
RG sense. While this turned out to be true, relying entirely
on tree-level power counting can be dangerous. Indeed, even a one-loop
analysis of Hertz's action would have revealed that the mean-field fixed
point is unstable, and the absence of such a calculation led to the instability
to not be noticed for 20 years.\cite{loops_footnote} Furthermore, integrating 
out the fermionic degrees of freedom obscures the fact that the problem 
contains two time scales, a diffusive one and a critical one, which in itself 
makes power counting very subtle. This, combined with the suggestive
relation between the ferromagnetic transition and the unidentified transition
discussed above, and the puzzling logarithmic discrepancies between the
critical behaviors found for the two transitions, makes it desirable to
have a theoretical description of the quantum ferromagnetic transition that
takes the form of a local field theory which facilitates a controlled loop
expansion and keeps the two time scales explicitly. Another motivation for 
constructing a local field theory is that
it will allow for a more explicit study of the effects of rare
regions\cite{us_rr} than was possible within the framework of
Refs.\ \onlinecite{us_dirty,us_clean}, although we will not pursue this
issue in the present paper.

It is the purpose of the present paper, and a second one to be referred
to as II,\cite{us_paper_II} to put these 
remaining questions to rest. We will focus on
the disordered case, although we expect analogous conclusions to hold for clean
systems.  By using a local field theory description, we will show that
Ref.\ \onlinecite{us_dirty} missed effects of the two time scales
that lead to logarithmic corrections to the Gaussian critical behavior.
Moreover, taking these effects into account leads to integral
equations for the relevant vertex functions that are identical to the
ones derived in Ref.\ \onlinecite{us_IFS}. The current formulation makes
it obvious that the transition described by these equations is the
quantum ferromagnetic one, and it allows to determine the exponents in
the ferromagnetic phase as well as those in the paramagnetic one. It
furthermore shows that the integral equations that were first derived in
Ref.\ \onlinecite{us_IFS} are exact, and it elucidates many physical points 
that were rather obscure in Ref.\ \onlinecite{us_dirty}, and to an even 
larger extent in Ref.\ \onlinecite{us_IFS}.

This paper is organized as follows. In Sec.\ \ref{sec:II} we use methods
developed in Ref.\ \onlinecite{us_fermions} to derive an
effective theory for disordered itinerant quantum ferromagnets that
systematically separates massive modes from soft ones, and explicitly
keeps all of the latter. In Sec.\ \ref{sec:III} we give a RG analysis
of this model. We first show how Hertz's fixed point, as well as the
Gaussian fixed point of Ref.\ \onlinecite{us_dirty}, emerge within this
framework. We then show that Hertz's fixed point is unstable, and the
Gaussian one marginally unstable, due to the existence of two separate
time scales, viz. a critical time scale associated with the order parameter
fluctuations, and a diffusive one associated with the additional particle-hole 
excitations. We identify an effective action that contains a stable
critical fixed point. This action is not Gaussian, and its
solution is therefore nontrivial and deferred to II. In Sec.\ \ref{sec:IV}
we discuss our results, in particular the relation of the present approach
to previous ones, and the complications that the presence of two time scales 
leads to in scaling considerations.

\section{Effective field theory}
\label{sec:II}

In this section we start with a simple model for interacting
electrons in a disordered environment. We then introduce the
ferromagnetic order parameter and identify all other soft
modes. Integrating out the massive modes leads to an
effective field theory that describes all of the soft modes
in the system. The general method employed here is the one
that was developed in Ref.\ \onlinecite{us_fermions}.

\subsection{Model of itinerant electrons}
\label{subsec:II.A}

Our starting point is a general field theoretic representation of the
partition function of a many-fermion system, which can be written in the 
form,\cite{NegeleOrland}
\begin{mathletters}
\label{eqs:2.1}
\begin{equation}
Z=\int D[{\bar\psi},\psi]\ \exp\left(S\left[{\bar\psi},\psi\right]\right)\quad.
\label{eq:2.1a}
\end{equation}
Here the functional integration measure is defined with respect to 
Grassmannian, or anticommuting, fields $\bar\psi$ and $\psi$, and $S$
is the action,
\begin{equation}
S = - \int_0^{\beta} d\tau \int d{\bf x}\ 
  {\bar\psi}_a({\bf x},\tau)\,{\partial\over\partial\tau}\,\psi_a({\bf x},\tau)
 - \int_0^{\beta} d\tau\ H(\tau)\ .
\label{eq:2.1b}
\end{equation}
\end{mathletters}%
We denote the spatial position by ${\bf x}$, and the imaginary time by $\tau$.
$H(\tau)$ is the Hamiltonian in imaginary time representation,
$\beta=1/T$ is the inverse temperature, $a=1,2$ denotes spin labels, and a
summation over repeated spin indices is implied. We choose
units such that $k_B = \hbar = e^2 = 1$. The Hamiltonian describes a fluid
of interacting electrons moving in a static random potential $v({\bf x})$,
\begin{mathletters}
\label{eqs:2.2}
\begin{eqnarray}
H(\tau) = \int d{\bf x}\ \left[{1\over 2m}\,\nabla {\bar\psi}_a({\bf x},\tau)
             \cdot\nabla\psi_a({\bf x}, \tau)\right.
\nonumber\\
             + [v({\bf x}) - \mu]\, {\bar\psi}_a({\bf x},\tau)\,
                               \psi_a({\bf x},\tau)\biggr]
\nonumber\\
          +\ {1\over 2} \int d{\bf x}\,d{\bf y}\ u({\bf x}-{\bf y})\ 
             {\bar\psi}_a({\bf x},\tau)\,{\bar\psi}_b({\bf y},\tau)\,
             \psi_b({\bf y},\tau)\,
\nonumber\\
             \times\psi_a({\bf x},\tau)\quad.
\label{eq:2.2a}
\end{eqnarray}
Here $m$ is the electron mass, $\mu$ is the chemical potential, and
$u({\bf x}-{\bf y})$ is the electron-electron interaction potential.
For simplicity, we assume that the random potential 
$v({\bf x})$ is delta-correlated
and obeys a Gaussian distribution $P[v({\bf x)}]$ with second moment
\begin{equation}
\left\{v({\bf x})\,v({\bf y})\right\}_{\rm dis} = {1\over 2\pi N_F 
   \tau_{\rm el}}\ \delta({\bf x} - {\bf y})\quad,
\label{eq:2.2b}
\end{equation}
where 
\begin{equation}
\{\ldots\}_{\rm dis} = \int D[v]\ P[v]\ (\ldots)\quad,
\label{eq:2.2c}
\end{equation}
\end{mathletters}%
denotes the disorder average, $N_F$ is the bare density of states per spin 
at the Fermi level, and $\tau_{\rm el}$ is the bare electron
elastic mean-free time. Our results will not be sensitive to the
simplifications inherent in the assumptions that lead to Eq.\ (\ref{eq:2.2b}).
We also mention that it would be possible to include more realistic features,
e.g. band structure, in the model. However, ultimately we will be
interested in universal behavior at a phase transition that is independent
of all microscopic details.
For our purposes it therefore is sufficient to study the model defined in
Eqs.\ (\ref{eqs:2.2}).\cite{Fermi_surface_footnote}

As in Ref.\ \onlinecite{us_dirty}, and as is standard practice in the
theory of magnetism, we
break the interaction part of the action $S$, which we denote by $S_{\rm int}$,
into spin-singlet and spin-triplet contributions, $S_{\rm }^{\,(s,t)}$. For
simplicity, we assume that the interactions are short-ranged in both of
these channels.\cite{short_range_footnote}
The spin-triplet interaction $S_{\rm int}^{\,(t)}$ describes interactions 
between spin-density fluctuations. This is the interaction
that causes ferromagnetism, and it therefore needs to be considered
separately. We thus write
\begin{equation}
S = S_0 + S_{\rm int}^{\,(t)}\quad,
\label{eq:2.3}
\end{equation}
with
\begin{mathletters}
\label{eqs:2.4}
\begin{equation}
S_{\rm int}^{\,(t)} = {\Gamma_t\over 2} \int d{\bf x}\,d\tau\ 
   {\bf n}_s({\bf x},\tau) \cdot {\bf n}_s({\bf x},\tau)\quad,
\label{eq:2.4a}
\end{equation}
where ${\bf n}_s$ is the electron spin-density vector with components,
\begin{equation}
n_s^i({\bf x},\tau) = {\bar\psi}_a({\bf x},\tau)\,
     \sigma_i^{ab}\,\psi_b({\bf x},\tau)\quad.
\label{eq:2.4b}
\end{equation}
Here the $\sigma_i$ $(i=1,2,3)$ are the Pauli matrices, and $\Gamma_t$ is the
spin-triplet interaction amplitude that is related to the interaction
potential $u$ in Eq.\ (\ref{eq:2.2a}) via
\begin{equation}
\Gamma_t = \frac{1}{2}\int d{\bf x}\ u({\bf x}) \quad.
\label{eq:2.4c}
\end{equation}
\end{mathletters}%
$S_0$ in Eq.\ (\ref{eq:2.3}) contains all other contributions to the
action. It reads explicitly, 
\begin{mathletters}
\label{eqs:2.5}
\begin{eqnarray}
S_0 = -\int_0^{\beta} d\tau \int d{\bf x}\ \biggl[
   {\bar\psi}_a({\bf x},\tau)\,{\partial\over\partial\tau}\,
                 \psi_a({\bf x},\tau)
\nonumber\\
   - {\bar\psi}_a({\bf x},\tau)\,\left(\frac{\nabla^2}{2m} + \mu\right)
             \psi_a({\bf x}, \tau)\biggr]
\nonumber\\
   - {\Gamma_s\over 2} \int_0^{\beta} d\tau \int d{\bf x}\ 
         n_c({\bf x},\tau)\,n_c({\bf x},\tau)
\nonumber\\
  -\int_0^{\beta} d\tau \int d{\bf x}\ v({\bf x})\,{\bar\psi}_a({\bf x},\tau)\,
     \psi_a({\bf x},\tau)\quad,
\label{eq:2.5a}
\end{eqnarray}
with $n_c$ the electron charge or number density,
\begin{equation}
n_c({\bf x},\tau) = {\bar\psi}_a({\bf x},\tau)\, \psi_a({\bf x},\tau)\quad,
\label{eq:2.5b}
\end{equation}
\end{mathletters}%
and $\Gamma_s$ the spin-singlet interaction amplitude.

Before we proceed, we integrate out the quenched disorder by means of
the replica trick.\cite{Grinstein} Performing the disorder average as
prescribed in Eq.\ (\ref{eq:2.2c}) replaces the last contribution to
the action $S_0$, Eq.\ (\ref{eq:2.5a}), by
\bea
S_{\rm dis}&=&\frac{1}{4\pi N_{\rm F}\tau_{\rm el}}\sum_{\alpha_1,
     \alpha_2=1}^{N}
  \int_0^{\beta} d\tau\,d\tau' \int d{\bf x}\ {\bar\psi}_a^{\alpha_1}
            ({\bf x},\tau)\, \nonumber\\
   &&\hskip 30pt\times \psi_a^{\alpha_1}({\bf x},\tau)\,
     {\bar\psi}_b^{\alpha_2}({\bf x},\tau')\,
     \psi_b^{\alpha_2}({\bf x},\tau')
   \quad,
\label{eq:2.6}
\eea
where $\alpha_1$ and $\alpha_2$ are replica indices, and $N\rightarrow 0$ is
the number of replicas. Of course, all other terms in the action also
are replicated $N$ times.

\subsection{Composite variables}
\label{subsec:II.B}

We now proceed by rewriting our model in terms of variables that are
more suitable for our purposes than the basic fermionic field. First
of all, we decouple the spin-triplet interaction by means of a
Hubbard-Stratonovich transformation. All other terms we rewrite in
terms of bosonic matrix fields $Q$ and $\wt\Lambda$. The latter
procedure exactly follows Ref.\ \onlinecite{us_fermions}, and we
refer the reader to that reference for details. Here we just mention
that $\wt\Lambda$ serves as a Lagrange multiplier whose physical
interpretation is a self energy, while $Q$ is 
isomorphic to bilinear products of fermion fields. We perform a Fourier
transform from imaginary time $\tau$ to Matsubara frequencies
$\omega_n = 2\pi T(n+1/2)$,
\bml
\label{eqs:2.7}
\bea
\psi_{n,a}({\bf x})&=&\sqrt{T}\int_0^{\beta}d\tau\ e^{i\omega_n\tau}\,
   \psi_a({\bf x},\tau)\quad,
\nonumber\\
{\bar\psi}_{n,a}({\bf x})&=&\sqrt{T}\int_0^{\beta}d\tau
   \ e^{-i\omega_n\tau}\,{\bar\psi}_a({\bf x},\tau)\quad.
\label{eq:2.7a}
\eea
For later reference, we also define a spatial Fourier transform
\bea
\psi_{n,a}({\bf k})&=&\frac{1}{\sqrt{V}}\int d{\bf x}\ e^{-i{\bf k}\cdot
   {\bf x}}\,\psi_{n,a}({\bf x})\quad,
\nonumber\\
{\bar\psi}_{n,a}({\bf k})&=&\frac{1}{\sqrt{V}}\int d{\bf x}\ 
   e^{i{\bf k}\cdot{\bf x}}\,{\bar\psi}_{n,a}({\bf x})\quad,
\label{eq:2.7b}
\eea
\eml%
and analogously for other position dependend quantities.
The isomorphism then takes the form,\cite{us_fermions}
\bea
Q_{12} &\cong&\ \frac{i}{2}\,\left( \begin{array}{cccc}
          -\psi_{1\uparrow}{\bar\psi}_{2\uparrow} &
             -\psi_{1\uparrow}{\bar\psi}_{2\downarrow} &
                 -\psi_{1\uparrow}\psi_{2\downarrow} &
                      \ \ \psi_{1\uparrow}\psi_{2\uparrow}  \\
          -\psi_{1\downarrow}{\bar\psi}_{2\uparrow} &
             -\psi_{1\downarrow}{\bar\psi}_{2\downarrow} &
                 -\psi_{1\downarrow}\psi_{2\downarrow} &
                      \ \ \psi_{1\downarrow}\psi_{2\uparrow}  \\
          \ \ {\bar\psi}_{1\downarrow}{\bar\psi}_{2\uparrow} &
             \ \ {\bar\psi}_{1\downarrow}{\bar\psi}_{2\downarrow} &
                 \ \ {\bar\psi}_{1\downarrow}\psi_{2\downarrow} &
                      -{\bar\psi}_{1\downarrow}\psi_{2\uparrow} \\
          -{\bar\psi}_{1\uparrow}{\bar\psi}_{2\uparrow} &
             -{\bar\psi}_{1\uparrow}{\bar\psi}_{2\downarrow} &
                 -{\bar\psi}_{1\uparrow}\psi_{2\downarrow} &
                      \ \ {\bar\psi}_{1\uparrow}\psi_{2\uparrow} \\
                    \end{array}\right)\ .
\nonumber\\
\label{eq:2.8}
\eea
Here all fields are understood to be taken at position ${\bf x}$,
and $1\equiv (n_1,\alpha_1)$, etc., comprises both frequency and
replica labels. It is convenient to expand the $4\times 4$ matrix 
in Eq.\ (\ref{eq:2.8}) in a spin-quaternion basis,
\be
Q_{12}({\bf x}) = \sum_{r,i=0}^3 (\tau_r\otimes s_i)\,{^i_rQ_{12}}({\bf x})
                 \quad
\label{eq:2.9}
\ee
and analogously for $\wt\Lambda$. Here
$\tau_0 = s_0 = \openone_2$ is the
$2\times 2$ unit matrix, and $\tau_j = -s_j = -i\sigma_j$, $(j=1,2,3)$,
with $\sigma_{1,2,3}$ the Pauli matrices. In this basis, $i=0$ and $i=1,2,3$
describe the spin-singlet and the spin-triplet, respectively. An explicit
calculation reveals that $r=0,3$ corresponds to the particle-hole channel
(i.e., products ${\bar\psi}\psi$), while $r=1,2$ describes the
particle-particle channel (i.e., products ${\bar\psi}{\bar\psi}$ or
$\psi\psi$). For our purposes the latter will not be of importance, and
we therefore drop $r=1,2$ from the spin-quaternion basis.

Denoting the Hubbard-Stratonovich field by $M$, we now can exactly
rewrite the partition function as an integral over the three fields
$Q$, $\wt\Lambda$, and $M$,
\bml
\label{eqs:2.10}
\be
Z = \int D[Q,{\wt\Lambda},M]\ e^{{\cal A}[Q,{\wt\Lambda},M]}\quad,
\label{eq:2.10a}
\ee
with an action
\bea
{\cal A}[Q,{\wt\Lambda},M]&=&{\cal A}_{\rm dis}[Q] 
   + {\cal A}_{\rm int}^{(s)}[Q] 
   + \frac{1}{2}\,\Tr\ln\left(G_0^{-1} - i{\wt\Lambda}\right)
\nonumber\\
   &&\hskip -30pt + \Tr\left({\wt\Lambda}Q\right)
   - \int d{\bf x}\sum_{\alpha}\sum_n\sum_{i=1}^{3} {^iM}_n^{\alpha}({\bf x})\,
          {^iM}_{-n}^{\alpha}({\bf x})
\nonumber\\
   &&\hskip -30pt  + \sqrt{2T\Gamma_t}\int d{\bf x}
                              \sum_{\alpha}\sum_n\sum_{i=1}^{3}
          {^iM}_n^{\alpha}({\bf x})\sum_{r=0,3} (\sqrt{-1})^r
\nonumber\\
   &&\times\sum_m \tr\left[\left(\tau_r\otimes s_i\right)\,
                   Q_{m,m+n}^{\alpha\alpha}({\bf x})\right]\ .
\label{eq:2.10b}
\eea
Here $\Tr$ denotes a trace over all degrees of freedom, including the
continuous real space position, while $\tr$ is a trace over all discrete
degrees of freedom that are not summed over explicitly. The first two
terms in Eq.\ (\ref{eq:2.10b}) read explicitly,
\be 
{\cal A}_{\rm dis}[Q] = \frac{1}{2\pi N_{\rm F}\tau_{\rm el}}\int d{\bf x}\ 
   \tr \left(Q({\bf x})\right)^2\quad,
\label{eq:2.10c}
\ee
\bea
{\cal A}_{\rm int}^{\,(s)}&=&\frac{T\Gamma^{(s)}}{2}\int d{\bf x}\sum_{r=0,3}
   (-1)^r \sum_{n_1,n_2,m}\sum_\alpha
\nonumber\\
&&\times\left[\tr \left((\tau_r\otimes s_0)\,Q_{n_1,n_1+m}^{\alpha\alpha}
({\bf x})\right)\right]
\nonumber\\
&&\times\left[\tr \left((\tau_r\otimes s_0)\,Q_{n_2+m,n_2}^{\alpha\alpha}
({\bf x})\right)\right]\quad,
\label{eq:2.10d}
\eea
They are simply the last two terms in Eq.\ (\ref{eq:2.5a}), rewritten in
terms of the $Q$. 
Finally, 
\be 
G_0^{\,-1} = -\partial_{\tau} +\nabla^2 /2m+\mu \quad,
\label{eq:2.10e}
\ee
is the inverse Green operator. For later reference we also note that if
$\wt\Lambda$ in the $\Tr\ln$ term in Eq.\ (\ref{eq:2.10b}) is replaced by
its saddle-point value, one obtains the inverse saddle-point Green function
$G_{\rm sp}$ as the argument of the logarithm.\cite{us_fermions} For our
purposes it suffices to treat the disorder contribution to the self energy
in Born approximation, and to neglect the (Hartree-Fock) interaction
contribution. $G_{\rm sp}$ then reads
\be
G_{\rm sp}({\bf k},\omega_n) \approx \left[i\omega_n - \frac{{\bf k}^2}{2m}
   + \mu + \frac{i}{2\tau_{\rm el}}\,\sgn \omega_n\right]^{-1}\quad.
\label{eq:2.10f}
\ee
\eml%
Notice that Eqs.\ (\ref{eqs:2.10}) can also be obtained
from Eqs.\ (2.22) - (2.25) in Ref.\ \onlinecite{us_fermions} by subjecting
the spin-triplet interaction term to a Gaussian transformation, and by
using the symmetry properties of the $Q$-matrices. However, the above
derivation makes it clear that the order parameter field $M$ is introduced
in a standard way, and the only difference to a standard treatment is that
the fermionic parts of the action have been
rewritten in terms of $Q$ and $\wt\Lambda$.

\subsection{Separation of soft and massive modes}
\label{subsec:II.C}

The reason for our rewriting of the fermionic part of the action in terms
of bosonic matrix fields in the previous subsection was that this formulation
is particularly well suited for a separation of soft and massive modes. For the
purpose of analyzing a Fermi liquid, this separation was carried out in
Ref.\ \onlinecite{us_fermions}, and we briefly recall the most important
points of that prodecure.\cite{WegnerSchafer} 
One first uses group theory arguments to show
that the most general $Q$ can be written in the form
\be
Q = {\cal S}\,P\,{\cal S}^{-1}\quad.
\label{eq:2.11}
\ee
Here $P$ is block-diagonal in Matsubara frequency space,
\be
P = \left( \begin{array}{cc}
       P^> & 0   \\
       0   & P^< \\
    \end{array} \right)\quad,
\label{eq:2.12}
\ee
with $P^>$ and $P^<$ matrices with elements $P_{nm}$ where $n,m>0$ and
$n,m<0$, respectively. For a system with $N$ replicas and $n$ Matsubara
frequencies, the matrices ${\cal S}$ are elements of the
homogeneous space ${\rm USp}(8Nn,{\cal C})/{\rm USp}(4Nn,{\cal C})\times 
{\rm USp}(4Nn,{\cal C})$.\cite{homogeneous_space_footnote}
As such they can be expressed in terms of
matrices $q$ whose elements $q_{nm}$ are restricted to frequency labels
$n\geq 0$, $m<0$,
\bml
\label{eqs:2.13}
\be
{\cal S} = \left( \begin{array}{cc}
                 \sqrt{1-bb^{\dagger}} & b   \\
                   -b^{\dagger}        & \sqrt{1-b^{\dagger} b} \\
           \end{array} \right)\quad,
\label{eq:2.13a}
\ee
where
\be
b(q,q^{\dagger}) = \frac{-1}{2}\,q\,f(q^{\dagger}q)\quad,
\label{eq:2.13b}
\ee
with
\be
f(x) = \sqrt{\frac{2}{x}}\,\left(1 - \sqrt{1 - x}\right)^{1/2}\quad.
\label{eq:2.13c}
\ee
\eml%
Ward identities ensure that the $P$ are massive while the $q$ are massless.
The latter are the diffusive modes or `diffusons' that we mentioned in the
Introduction. Similarly, if $\wt\Lambda$ is transformed according to
\be
\Lambda ({\bf x}) = {\cal S}^{-1}({\bf x})\,{\tilde\Lambda}({\bf x})\,{\cal S}
                        ({\bf x})\quad,
\label{eq:2.14}
\ee
$\Lambda$ can be shown to be massive. This representation thus
achieves the desired separation of modes. In Ref.\ \onlinecite{us_fermions}
the $q$ were the only soft modes. In the present case, the order parameter
field $M$ is massive in the paramagnetic phase, but becomes soft at
criticality, so it must be handled together with the diffusons.

The next step is to expand the massive modes about their expectation
values,
\be
P = \langle P\rangle + \Delta P\quad,\quad \Lambda = \langle\Lambda\rangle
                                            + \Delta\Lambda\quad.
\label{eq:2.15}
\ee
As explained in Ref.\ \onlinecite{us_fermions}, the expectation values
$\langle P\rangle$ and $\langle\Lambda\rangle$ can be replaced by simple
saddle-point approximations. The saddle point used in this reference is
also a saddle point of our current action, Eq.\ (\ref{eq:2.10b}), if it
is supplemented by the saddle-point value of $M$, $\langle M\rangle = 0$.
If $P$ and $\Lambda$ are integrated out in saddle-point approximation,
i.e. if one neglects all fluctuations of these fields, then one obtains
for the fermionic degrees of freedom the nonlinear sigma model (NL$\sigma$M)
that was first proposed by Wegner\cite{Wegner}
as an effective field theory for the
disordered electron problem, and later studied extensively by him and
others.\cite{us_R,NLsigmaM_footnote} Following the same procedure here, 
we get the
NL$\sigma$M, but without the spin-triplet interaction term, from the
first four terms on the r.h.s. of Eq.\ (\ref{eq:2.10b}), and a coupling
between the order parameter field and both the soft and the massive
fermionic modes from the last one. We will also need the corrections to
the NL$\sigma$M, and thus rewrite the action in the form
\bea
{\cal A}[q,M,\Delta P,\Delta\Lambda]&=&{\cal A}_{{\rm NL}\sigma{\rm M}}[q]
   + \delta{\cal A}[\Delta P,\Delta\Lambda,q]
\nonumber\\
 &&\hskip - 40pt - \int d{\bf x}\sum_{\alpha}\sum_n\sum_{i=1}^{3} 
     {^iM}_n^{\alpha}({\bf x})\, {^iM}_{-n}^{\alpha}({\bf x})
\nonumber\\
 &&\hskip -60pt  + \sqrt{\pi TK_t}\int d{\bf x}
                              \sum_{\alpha}\sum_n\sum_{i=1}^{3}
          {^iM}_n^{\alpha}({\bf x})\sum_{r=0,3} (\sqrt{-1})^r
\nonumber\\
   &&\hskip -50pt \times\sum_m \tr\left(\tau_r\otimes s_i\right)\,
      \left[{\hat Q}^{\alpha\alpha}_{m,m+n}({\bf x}) \right.
\nonumber\\
       &&\hskip -30pt + \left.\frac{2}{\pi N_{\rm F}}\,
      \left({\cal S}\Delta P{\cal S}^{-1}\right)
              ^{\alpha\alpha}_{m,m+n}({\bf x})\right]\quad.
\label{eq:2.16}
\eea
Here $K_t = \pi N_{\rm F}^2\Gamma_t/2$. 
${\cal A}_{{\rm NL}\sigma{\rm M}}$ is the action of the nonlinear
sigma model,
\bml
\label{eqs:2.17}
\bea
{\cal A}_{{\rm NL}\sigma{\rm M}}&=&\frac{-1}{2G}\int d{\bf x}\
     \tr\left(\nabla\hat Q ({\bf x})\right)^2
\nonumber\\
&&\hskip -50pt + 2H \int d{\bf x}\ \tr\left(\Omega\,{\hat Q}({\bf x})
     \right)
 + {\cal A}_{\rm int}^{(s)}[\pi\,N_{\rm F}{\hat Q}/2]\quad,
\label{eq:2.17a}
\eea
with ${\cal A}_{\rm int}^{(s)}$ from Eq.\ (\ref{eq:2.10d}), and
\be
{\hat Q} = \left( \begin{array}{cc}
                 \sqrt{1-qq^{\dagger}} & q   \\
                    q^{\dagger}        & -\sqrt{1-q^{\dagger} q} \\
           \end{array} \right)\quad,
\label{eq:2.17b}
\ee
in terms of $q$, and $\Omega$ a frequency matrix with elements
\be
\Omega_{12} = \left(\tau_0\otimes s_0\right)\,\delta_{12}\,\omega_{n_1}\quad,
\label{eq:2.17c}
\ee
\eml%
The coupling constants $G$ and $H$ are proportional to the inverse
conductivity, $G\propto 1/\sigma$, and the specific heat coefficient,
$H\propto\gamma\equiv\lim_{T\rightarrow 0}\,C_V/T$,
respectively.\cite{us_R,us_fermions,CC}

$\delta{\cal A}$ contains the corrections to the nonlinear sigma model
that were given in Ref.\ \onlinecite{us_fermions}. We list explicitly
the terms that are bilinear in the massive fluctuations $\Delta P$ and
$\Delta\Lambda$, but do not contain couplings between
the massive modes and $q$,
\bea
\delta{\cal A}^{(2)}&=&{\cal A}_{\rm dis}[\Delta P]
   + \int d{\bf x}\ \tr\,\left(\Delta\Lambda ({\bf x})\,
     \Delta P ({\bf x})\right)
\nonumber\\
   &&\hskip -30pt + \frac{1}{4} \int d{\bf x}\,d{\bf y}\ 
      \tr\big(G({\bf x}-{\bf y})\,
   \Delta\Lambda({\bf y})\,G({\bf y}-{\bf x})\,\Delta\Lambda ({\bf x})\bigr)
     \quad.
\nonumber\\
\label{eq:2.18}
\eea

\subsection{Effective field theory for the soft modes}
\label{subsec:II.D}

So far we have exactly rewritten the microscopic action in a form that 
separates the soft modes from the massive ones. In this subsection we 
approximately integrate out the massive modes to arrive at an effective 
action that is capable of describing
the critical behavior at the ferromagnetic transition. We will also add some
terms to the bare action that we will find later to be generated by the
renormalization group. In Sec.\ \ref{sec:III} we will derive these terms, and
will also justify our approximations and show that they do not influence the
asymptotic critical behavior.

\subsubsection{Integrating out the massive modes}
\label{subsubsec:II.D.1}

We now need to dispose of the massive modes. Clearly, we cannot just
ignore the massive fluctuations. It is obvious from 
Eqs.\ (\ref{eq:2.16}), (\ref{eqs:2.17}) that they are needed to bring
the purely $M$-dependent part of the action in a standard 
Landau-Ginzburg-Wilson (LGW) form. Let us first integrate out $\Delta P$ and
$\Delta\Lambda$ in a Gaussian approximation, while neglecting the coupling
between $q$ and these massive fluctuations. We will consider the effects
of this coupling later. From Eq.\ (\ref{eq:2.18}) and the last term in
Eq.\ (\ref{eq:2.16}) with ${\cal S}=1$ we obtain a contribution to the
action that is quadratic in the order parameter field $M$. Combining it
with the $M^2$ term in Eq.\ (\ref{eq:2.16}) yields a term proportional to
\bml
\label{eqs:2.19}
\be
\sum_{\bf k}\sum_{\alpha}\sum_n\sum_{i=1}^3 {^iM}_n^{\alpha}({\bf k})\,
   \left[1 + 2\Gamma_t{\tilde\chi}({\bf k},\Omega_n)\right]\,
       {^iM}_{-n}^{\alpha}(-{\bf k})\quad,
\label{eq:2.19a}
\ee
where
\be
{\tilde\chi}({\bf k},\Omega_n) = T\sum_{n_1,n_2}\,\Theta(n_1n_2)\,
   \delta_{n_1-n_2,n}\,{\cal D}_{n_1n_2}({\bf k})\quad,
\label{eq:2.19b}
\ee
is a restricted spin susceptibility that is given in terms of
\be
{\cal D}_{nm}({\bf k}) = \varphi_{nm}({\bf k})\,
   \left[1 - \frac{1}{2\pi N_{\rm F}
   \tau_{\rm el}}\,\varphi_{nm}({\bf k})\right]^{-1}
\label{eq:2.19c}
\ee
with
\be
\varphi_{nm}({\bf k}) = \frac{1}{V}\sum_{\bf p} G_{\rm sp}({\bf p},\omega_n)\,
   G_{\rm sp}({\bf p}+{\bf k},\omega_m)\quad.
\label{eq:2.19d}
\ee
\eml%
Here $G_{\rm sp}$ is the saddle-point Green function from
Eq.\ (\ref{eq:2.10f}). The Theta-function in Eq.\ (\ref{eq:2.19b}),
which restricts the frequency sum to frequencies that both have the same sign,
results in $\tilde\chi$ being the non-hydrodynamic part of the spin 
susceptibility of noninteracting disordered electrons. For small frequencies 
and wavenumbers, it reads
\bml
\label{eqs:2.20}
\be
{\tilde\chi}({\bf k},\Omega_n) = -N_{\rm F} + O({\bf k}^2,\Omega_n)\quad.
\label{eq:2.20a}
\ee
For later reference, we also note that for $nm<0$, ${\cal D}_{nm}$ is the
basic diffusion propagator. In the limit of small frequencies and
wavenumbers one finds
\bea
{\cal D}_{nm}({\bf k}) &=&\frac{2\pi N_{\rm F}}{D{\bf k}^2 
    + \vert\Omega_{n-m}\vert}\qquad (nm<0)\quad,
\nonumber\\
&\equiv&2\pi N_{\rm F}GH\,{\cal D}_{n-m}({\bf k})\quad,
\label{eq:2.20b}
\eea
with
\be
{\cal D}_n({\bf k}) = \frac{1}{{\bf k}^2 + GH\vert\Omega_{n}\vert}\quad.
\label{eq:2.20c}
\ee
\eml%
Here $D=1/GH$ is a bare diffusion coefficient, and $\Omega_n=2\pi Tn$ 
is a bosonic Matsubara frequency.

Within the approximation, Eq.\ (\ref{eq:2.20a}), the effective action has 
the form
\bea
{\wt{\cal A}}_{\rm eff}&=&-\sum_{\bf k}\sum_{\alpha}\sum_n\sum_{i=1}^3
   {^iM}_n({\bf k})\,\left[t + a{\bf k}^2 + b\vert\Omega_n\vert\right]
\nonumber\\
&&\hskip 100pt \times{^iM}_{-n}(-{\bf k}) 
            + {\cal A}_{{\rm NL}\sigma{\rm M}}[{\hat Q}]
\nonumber\\
&&\hskip -20pt  + \sqrt{\pi TK_t}\int d{\bf x}
                              \sum_{\alpha}\sum_n\sum_{i=1}^{3}
          {^iM}_n^{\alpha}({\bf x})\sum_{r=0,3} (\sqrt{-1})^r \sum_m
\nonumber\\
&&\times\tr\left[\left(\tau_r\otimes s_i\right)\,
      {\hat Q}^{\alpha\alpha}_{m,m+n}({\bf x}) \right]\quad.
\label{eq:2.21}
\eea
Here $t=1-2N_{\rm F}\Gamma_t$, $a$ and $b$ are constants, 
and in the first term we have truncated the
gradient expansion after the leading wavenumber and frequency dependent terms.
Notice that the first term taken in isolation describes a ferromagnetic
transition at $t=0$, which represents the Stoner criterion.

The following structure emerges. Our effective action takes the form of a LGW
action for the order parameter field $M$, a nonlinear sigma model for the
soft fermionic modes, and a coupling between the order parameter and the
composite fermion field ${\hat Q}$. In our current approximation, only the
Gaussian part of the LGW action appears. However, keeping terms of higher order
in $\Delta P$ would clearly produce terms of higher order in $M$, with 
coefficients that are numbers in the limit of zero wavenumbers and frequencies.
For our purposes it will suffice to keep the term of order $M^4$. We thus 
should add to the action a term
\bea
{\cal A}_{\rm LGW}^{(4,1)}&=&u_4\int\hskip -2pt d{\bf x}\ T\hskip -5pt
   \sum_{n_1,n_2,n_3}\hskip -2pt \sum_{\alpha}
   \left({\bf M}^{\alpha}({\bf x},\omega_{n_1})
        \cdot {\bf M}^{\alpha}({\bf x},\omega_{n_2})\right)
\nonumber\\
&&\times   \left({\bf M}^{\alpha}({\bf x},\omega_{n_3})\cdot
     {\bf M}^{\alpha}({\bf x},-\omega_{n_1}-\omega_{n_2}-\omega_{n_3})\right)
      \quad,
\nonumber\\
\label{eq:2.22}
\eea
with $u_4$ a number, and ${\bf M}$ the 3-vector whose components are the $^iM$.
A detailed derivation shows that there is also a term of third order 
in ${\bf M}$, with a 
coefficient that is proportional to either a gradient or a time derivative. 
This term is irrelevant for the critical behavior at the ferromagnetic 
transition,\cite{us_dirty} and we neglect it.

\begin{figure}[t,h]
\centerline{\psfig{figure=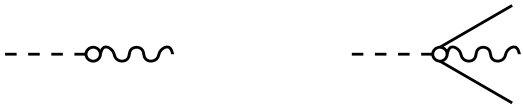,width=70mm}\vspace*{5mm}}
\caption{$M$-$\Delta P$, and $M$-$\Delta P$-$q^2$ vertices. Broken lines
 denote $M$, solid lines $q$, and wiggly lines $\Delta P$.}
\label{fig:1}
\end{figure}

\begin{figure}[t,h]
\centerline{\psfig{figure=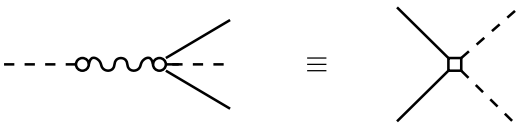,width=70mm}\vspace*{5mm}}
\caption{Effective $M^2$-$q^2$ vertex obtained by contracting the $\Delta P$
 fluctuations of the two vertices shown in Fig.\ \ref{fig:1}.}
\label{fig:2}
\end{figure}

\begin{figure}[t,h]
\centerline{\psfig{figure=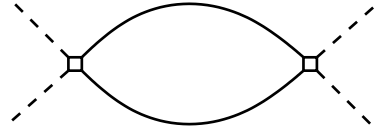,width=50mm}\vspace*{5mm}}
\caption{One-loop renormalization of the $M^4$ vertex by means of the
 vertex in Fig.\ \ref{fig:2}.}
\label{fig:3}
\end{figure}

Both from simple physical considerations, and from Ref.\ \onlinecite{us_dirty},
it is obvious that another term of $O(M^4)$ must exist. This is the 
`random mass'
term that reflects spatial fluctuations in the location of the critical point,
and it has the structure
\bea
{\cal A}_{\rm LGW}^{(4,2)}&=&v_4\int d{\bf x}\sum_{n_1,n_2}\sum_{\alpha,\beta}
   \left\vert{\bf M}^{\alpha}({\bf x},\omega_{n_1})\right\vert^2
\nonumber\\
&&\hskip 60pt \times   \left\vert{\bf M}^{\beta}({\bf x},\omega_{n_2})\right
    \vert^2 \quad,
\label{eq:2.23}
\eea
with $v_4$ a number. To see how this term arises in our present formulation, we
go back to Eq.\ (\ref{eq:2.16}). If we expand the ${\cal S}$ in the last term 
in powers of $q$, the two lowest order 
contributions\cite{cal_S_footnote,MDPq_footnote} have the structures
\bml
\label{eqs:2.24}
\bea
\sqrt{T}\int d{\bf x}\ M({\bf x})\,\Delta P({\bf x})\quad,
\label{eq:2.24a}\\
\sqrt{T}\int d{\bf x}\ M({\bf x})\,\Delta P({\bf x})\,q({\bf x})\,q({\bf x})
   \quad.
\label{eq:2.24b}
\eea
Here we have dropped frequency labels and all other degrees of freedom that are
irrelevant for power counting purposes. Let us depict the fields $M$, 
$\Delta P$, 
and $q$ by dashed, wavy, and solid lines, respectively. Then the two vertices
in Eqs.\ (\ref{eq:2.24a},\ref{eq:2.24b}) 
have the structures shown in Fig.\ \ref{fig:1}.
Contracting the wavy lines yields an effective vertex of the form
\be
T\int d{\bf x}\ M^2({\bf x})\,q^2({\bf x})\quad,
\label{eq:2.24c}
\ee
\eml%
which is shown in Fig.\ \ref{fig:2}. 
This vertex can be used to construct the
one-loop contribution to the term of $O(M^4)$ that is shown in 
Fig.\ \ref{fig:3}.
Calculating the diagram yields Eq.\ (\ref{eq:2.23}). We will further
discuss this term from a power-counting point of view in
Sec.\ \ref{subsubsec:III.C.2} below.

\subsubsection{Effective action, and leading corrections}
\label{subsubsec:II.D.2}

We now can assemble our effective action by collecting the various terms
derived in the previous subsection. 
Before doing so, however, it is illustrative
and convenient to add a term to the LGW part of the action that will be
generated by the RG in Sec.\ \ref{sec:III}. As we will see, for dimensions
$2<d<4$ the gradient squared term in the first contribution to $\tilde{\cal A}$
is {\em not} the leading wavenumber dependence. Rather, as one would expect
from Ref.\ \onlinecite{us_dirty}, there is a term proportional to 
$\vert{\bf k}\vert^{d-2}$, which first appears at one-loop order. We therefore
add this term right away. Notice that the resulting action is still
local in the sense of Ref.\ \onlinecite{local_footnote}.
It will also turn out that the 
term in the LGW action that is linear in frequency is not 
the leading frequency dependence. Rather, the
coupling between $M$ and ${\hat Q}$ effectively produces a term proportional to
$\vert\Omega_n\vert/{\bf k}^2$.\cite{Hertz} 
Physically, this term reflects the fact that
spin is a conserved variable, i.e. the characteristic frequency scale vanishes
in the long-wavelength limit even away from the critical point.
We can therefore drop the frequency dependence
from the first term in Eq.\ (\ref{eq:2.21}).

Taking all of these points into account, we obtain the following effective
action,
\bml
\label{eqs:2.25}
\be
{\cal A}_{\rm eff} = {\cal A}_{\rm LGW}[M] 
                     + {\cal A}_{{\rm NL}\sigma{\rm M}}[q]
                     + {\cal A}_{\rm c}[M,q]\quad.
\label{eq:2.25a}
\ee
Here ${\cal A}_{\rm LGW}$ is the modified LGW part of the action,
\bea
{\cal A}_{\rm LGW}[M]&=&-\sum_{\bf k}\sum_{\alpha}\sum_n\sum_{i=1}^3
   {^iM}_n^{\alpha}({\bf k})\,\left[t + a_{d-2}\vert{\bf k}\vert^{d-2} \right.
\nonumber\\
&&\left.\hskip 60pt + a_2{\bf k}^2\right]\,{^iM}_{-n}^{\alpha}(-{\bf k})
\nonumber\\
&&\hskip -30pt + {\cal A}_{\rm LGW}^{(4,1)}[M]
   + {\cal A}_{\rm LGW}^{(4,2)}[M]
\label{eq:2.25b}
\eea
with $a_{d-2}$ and $a_2$ constants, and ${\cal A}_{\rm LGW}^{(4,1)}$ and
${\cal A}_{\rm LGW}^{(4,2)}$ from Eqs.\ (\ref{eq:2.22}) and (\ref{eq:2.23}),
respectively. The nonlinear sigma model part of the action, 
${\cal A}_{{\rm NL}\sigma{\rm M}}$, has been given in Eq.\ (\ref{eq:2.17a}),
and ${\cal A}_{\rm c}$ represents the coupling between $M$ and $q$,
\bea
{\cal A}_{\rm c}[M,q]&=&\sqrt{\pi TK_t}\int \hskip -2pt d{\bf x}
                              \sum_{\alpha}\sum_n\sum_{i=1}^{3}
         {^iM}_n^{\alpha}({\bf x})\hskip -2pt \sum_{r=0,3} (\sqrt{-1})^r
\nonumber\\
&&\times\sum_m\tr\left[\left(\tau_r\otimes s_i\right)\,
      {\hat Q}^{\alpha\alpha}_{m,m+n}({\bf x}) \right]\quad.
\label{eq:2.25c}
\eea
\eml%
It is useful to define a field
\bml
\label{eqs:2.26}
\be
b_{12}({\bf x}) = \sum_{i,r} (\tau_r\otimes s_i)\,{^i_rb}_{12}({\bf x})\quad,
\label{eq:2.26a}
\ee
with components
\bea
{^i_rb}_{12}({\bf k})&=&\delta_{\alpha_1\alpha_2} (-)^{r/2}\sum_n
     \delta_{n,n_1-n_2}\left[{^iM}_n^{\alpha_1}({\bf k})\right.
\nonumber\\
 &&\hskip 50pt \left.+ (-)^{r+1}\,{^iM}_{-n}^{\alpha_1}({\bf k})\right]\quad.
\label{eq:2.26b}
\eea
\eml%
In terms of $b$, the coupling part of the action can be written
\be
{\cal A}_{\rm c}[b,q] = -\frac{1}{2}\,\sqrt{\pi TK_t}\int d{\bf x}\ \tr
   \left(b({\bf x})\,{\hat Q}({\bf x})\right)\quad.
\label{eq:2.27}
\ee

We stress that we have added the $\vert{\bf k}\vert^{d-2}$ term in
Eq.\ (\ref{eq:2.25b}) for convenience only. One could equally well
work with the theory given by our action with $a_{d-2}=0$, 
but some effects that are included in the bare ${\cal A}_{\rm eff}$ would
then appear only at loop level. We also stress that the coefficients of
all non-Gaussian terms in ${\cal A}_{\rm eff}$ are finite in the limit
of zero wavenumbers and frequencies, so the theory is local in the sense
of Ref.\ \onlinecite{local_footnote}, in contrast to the situation in
Ref.\ \onlinecite{us_dirty}. More importantly, the
physically relevant point is that the current formulation, as opposed
to the one in Ref.\ \onlinecite{us_dirty}, makes obvious the existence of two
time scales, as we will see in Sec.\ \ref{sec:III} below.

In order to justify the omission of the terms left out of our effective action,
we also need to know the structure of the omitted terms. We therefore list 
these here, using notation that leaves out anything that is not needed for 
power counting purposes.

First of all we have the corrections to the nonlinear sigma model. In addition
to Eq.\ (\ref{eq:2.18}), they consist of those terms in the $\Tr\ln$ term in
Eq.\ (\ref{eq:2.10b}) that contain cubic or higher orders of $\Delta\Lambda$,
the part of ${\cal A}_{\rm int}^{(s)}$ that contains the massive fluctuation
$\Delta P$, and corrections to the saddle-point approximation for 
$\langle P\rangle$ in Eq.\ (\ref{eq:2.15}). These have all been discussed in
Ref.\ \onlinecite{us_fermions}, and the same discussion applies here.

Second, there are the terms that couple $M$, $\Delta P$, and $q$, see the
second term in the bracket in the last line of Eq.\ (\ref{eq:2.16}). In
general, they have the structure
\be
d_n\,\sqrt{T} \int d{\bf x}\,M({\bf x})\,\Delta P({\bf x})\,q^n({\bf x})\quad,
\label{eq:2.28}
\ee
with coupling constants $d_n$ and $n=0,2,3,\ldots$.\cite{MDPq_footnote} 
The first two terms in this expansion have already been
given in Eqs.\ (\ref{eqs:2.24}).

\subsubsection{Observables}
\label{subsubsec:II.D.3}

For a physical interpretation of any results obtained from our effective
action we need to identify the appropriate observables in terms of the
coupling constants of the theory. It is obvious, and easily confirmed by
keeping a source term for the electron spin-density, that the expectation
value of the order parameter field $M$ determines the magnetization $m$.
Specifically,
\bml
\label{eqs:2.29}
\be
m = \mu_{\rm B}\,\sqrt{2T/\Gamma_t}\,\langle{^iM}_{n=0}^{\alpha}({\bf x})
   \rangle\quad.
\label{eq:2.29a}
\ee
The two-point $M$-vertex is therefore proportional to the magnetic
susceptibility. 

Other relevant observables are the specific heat coefficient
\be
\gamma = \frac{4\pi}{3}\,H\quad,
\label{eq:2.29b}
\ee
and the electrical conductivity
\be
\sigma = \frac{8}{\pi G}\quad,
\label{eq:2.29c}
\ee
(see Sec.\ \ref{subsec:II.C}). Also of interest is the electronic
density of states per spin\cite{us_fermions}
\be
N(\epsilon_{\rm F}+\omega) = N_{\rm F}\,\langle
   {^0_0{\hat Q}}_{nn}^{\alpha\alpha}({\bf x})\rangle_{i\omega_n\rightarrow
      \omega + i0}\quad,
\label{eq:2.29d}
\ee
\eml%
where the energy or frequency $\omega$ is measured from the Fermi
energy $\epsilon_{\rm F}$.

\section{Renormalization group analysis}
\label{sec:III}

In this section, we first consider low-order perturbation theory to
see how those terms in Eqs.\ (\ref{eqs:2.25}) that
were not in the bare action are generated. We then do a power-counting
analysis to determine the minimal effective action that needs to be analyzed
in order to find the critical behavior at the ferromagnetic transition.
Finally, we show that the terms that were omitted from the effective
action are irrelevant, in the RG sense, for the critical behavior.

\subsection{Perturbation theory}
\label{subsec:III.A}

We first set up a standard perturbative expansion for our effective action,
starting with the Gaussian theory.

\subsubsection{Gaussian propagators}
\label{subsubsec:III.A.1}

We start by expanding the effective action, Eqs.\ (\ref{eqs:2.25}), to bilinear
order in $M$ (or $b$) and $q$. We obtain for the Gaussian action
\bml
\label{eqs:3.1}
\bea
{\cal A}_{\rm G}[M,q]&=&-\sum_{\bf k}\sum_n\sum_{\alpha}\sum_{i=1}^3
   {^iM}_n^{\alpha}({\bf k})\,u_2({\bf k})\,{^iM}_{-n}^{\alpha}(-{\bf k})
\nonumber\\
&&- \frac{4}{G}\sum_{\bf k}\sum_{1,2,3,4}\sum_{i,r}{^i_rq}_{12}({\bf k})\,
        {^i\Gamma}_{12,34}^{(2)}({\bf k})\,{^i_rq}_{34}(-{\bf k})
\nonumber\\
&&+ 4\sqrt{\pi TK_t}\sum_{\bf k}\sum_{12}\sum_{i,r}{^i_rq}_{12}({\bf k})\,
    {^i_rb}_{12} (-{\bf k})\quad,
\nonumber\\
\label{eq:3.1a}
\eea
where
\be
u_2({\bf k}) = t + a_{d-2}\vert{\bf k}\vert^{d-2} + a_2{\bf k}^2\quad.
\label{eq:3.1b}
\ee
The bare two-point $q$-vertex reads
\bea
{^0\Gamma}_{12,34}^{(2)}({\bf k})&=&\delta_{13}\delta_{24}\left({\bf k}^2
   + GH\Omega_{n_1-n_2}\right) 
\nonumber\\
&&\hskip -3pt + \delta_{1-2,3-4}\delta_{\alpha_1\alpha_2}
     \delta_{\alpha_1\alpha_3}\,2\pi TGK_s\quad,
\label{eq:3.1c}\\
{^{1,2,3}\Gamma}_{12,34}^{(2)}({\bf k})&=&\delta_{13}\delta_{24}\,\left(
   {\bf k}^2 + GH\Omega_{n_1-n_2}\right)\quad,
\label{eq:3.1d}
\eea
\eml%
with $K_s = -\pi\,N_{\rm F}^2\,\Gamma_s/2$. 

The quadratic form defined by this Gaussian action is easily inverted. For
the order parameter correlations we find
\bml
\label{eqs:3.2}
\bea
\langle{^iM}_n^{\alpha}({\bf k})\,{^jM}_m^{\beta}({\bf p})\rangle&=&
       \delta_{{\bf k},
   -{\bf p}}\,\delta_{n,-m}\,\delta_{ij}\,\delta_{\alpha\beta}\,\frac{1}{2}\,
    {\cal M}_n({\bf k})\quad,
\nonumber\\
\label{eq:3.2a}\\
\langle{^i_rb}_{12}({\bf k})\,{^j_sb}_{34}({\bf p})\rangle&=&
   -\delta_{{\bf k},-{\bf p}}\,\delta_{rs}\,\delta_{ij}\,
   \delta_{\alpha_1\alpha_2}\,\delta_{\alpha_1\alpha_3}
\nonumber\\
&&\hskip -35pt\times {\cal M}_{n_1-n_2}({\bf k})\,\left[\delta_{1-2,3-4}
        - (-)^r\delta_{1-2,4-3}\right]\quad,
\nonumber\\
\label{eq:3.2b}
\eea
in terms of the paramagnon propagator
\be
{\cal M}_n({\bf k}) = \frac{1}{t + a_{d-2}\vert{\bf k}\vert^{d-2} 
       + a_2\,{\bf k}^2
   + \frac{GK_t\vert\Omega_n\vert}{{\bf k}^2 + GH\vert\Omega_n\vert}}\quad.
\label{eq:3.2c}
\ee
\eml%
Notice that the coupling between the order parameter field and the
fermionic degrees of freedom has produced the dynamical piece of ${\cal M}$
that is characteristic of disordered itinerant ferromagnets.\cite{Hertz}

For the fermionic propagators we find
\bml
\label{eqs:3.3}
\be
\langle{^i_rq}_{12}({\bf k})\,{^j_sq}_{34}({\bf p})\rangle =
   \delta_{{\bf k},-{\bf p}}\,\delta_{rs}\,\delta_{ij}\,\frac{G}{8}\,
   {^i\Gamma}_{12,34}^{(2)\,-1}({\bf k})\quad,
\label{eq:3.3a}
\ee
in terms of the inverse of Eq.\ (\ref{eq:3.1c}),
\bea
{^0\Gamma}_{12,34}^{(2)\,-1}({\bf k})&=&\delta_{13}\,\delta_{24}\,
   {\cal D}_{n_1-n_2}
   ({\bf k}) - \delta_{1-2,3-4}\,\delta_{\alpha_1\alpha_2}\,
      \delta_{\alpha_1\alpha_3}\,
\nonumber\\
&&\times 2\pi TGK_s\,{\cal D}_{n_1-n_2}({\bf k})\,
       {\cal D}^{(\rm s)}_{n_1-n_2}({\bf k})\ ,
\label{eq:3.3b}
\eea
and
\bea
{^{1,2,3}\Gamma}_{12,34}^{(2)\,-1}({\bf k})&=&\delta_{13}\,\delta_{24}\,
   {\cal D}_{n_1-n_2}({\bf k}) - \delta_{1-2,3-4}\,\delta_{\alpha_1\alpha_2}
      \delta_{\alpha_1\alpha_3}\,
\nonumber\\
&&\times 2\pi TGK_t\,\left({\cal D}_{n_1-n_2}({\bf k})
        \right)^2\,{\cal M}_{n_1-n_2}({\bf k})\ .
\nonumber\\
\label{eq:3.3c}
\eea
Here ${\cal D}^{(\rm s)}$ is the spin-singlet propagator, which in
the limit of long wavelengths and small frequencies reads\cite{us_R}
\be
{\cal D}_{n}^{(\rm s)}({\bf k}) = \frac{1}{{\bf k}^2 + G(H+K_s)\Omega_n}\quad.
\label{eq:3.3d}
\ee
\eml%

Finally, due to the coupling between $M$ and $q$ we have a mixed propagator
\bea
\langle{^i_rq}_{12}({\bf k})\,{^j_sb}_{34}({\bf p})\rangle&=&
   -\delta_{{\bf k},-{\bf p}}\,\delta_{rs}\,\delta_{ij}\,
   \delta_{\alpha_1\alpha_2}\,\delta_{\alpha_1\alpha_3}\,\frac{G}{2}\,
\sqrt{\pi TK_t}\,
\nonumber\\
&&\hskip -70pt {\cal D}_{n_1-n_2}({\bf k})\,{\cal M}_{n_1-n_2}({\bf k})\,
   \left[\delta_{1-2,3-4} + (-)^{r+1}\delta_{1-2,4-3}\right] \, .
\nonumber\\
\label{eq:3.4}
\eea

\subsubsection{One-loop order}
\label{subsubsec:III.A.2}

\begin{figure}[t]
\centerline{\psfig{figure=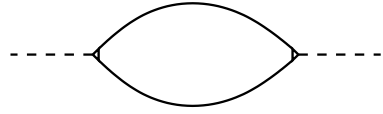,width=50mm}\vspace*{5mm}}
\caption{One-loop renormalization of the vertex $u_2$.}
\label{fig:4}
\end{figure}

Let us now consider the renormalization of the $M^2$-vertex $u_2$,
Eq.\ (\ref{eq:3.1b}). At one-loop order, the relevant diagram is the one
shown in Fig.\ \ref{fig:4}. While the complete result is rather
involved, it can be simplified by means of the following observation.
The structure of the diagram leads to a frequency-momentum integral
over diffusion poles multiplied by one or more paramagnon propagators.
Inspection shows that the frequency in this integral
scales like a wavenumber squared. To leading order
in the distance from the critical point, and for the purpose of
obtaining leading infrared singularities, we therefore can use the
following approximation in the integrand,
\bml
\label{eqs:3.5}
\be
G\,K_t\,\Omega_l\,{\cal D}_l({\bf p})\,{\cal M}_l({\bf p}) =
   1 + O(t,\vert{\bf p}\vert^d)\quad.
\label{eq:3.5a}
\ee
At zero external frequency the diagram then yields
\be
\delta u_2({\bf k}) = -\frac{G^2}{2}\,\frac{K_t}{V} \sum_{\bf p} 2\pi T
   \sum_{l=0}^{\infty}
   {\cal D}_l({\bf p})\,{\cal D}_l({\bf p}+{\bf k})\ .
\label{eq:3.5b}
\ee
\eml%
Performing the integrals shows that this diagram provides a 
finite renormalization
of the coefficient $a_{d-2}$ of the $\vert{\bf k}\vert^{d-2}$ term. In
particular, it would generate this term if it were not present in the
bare action. This is the reason why we added this term by hand in
Sec.\ \ref{subsubsec:II.D.2}.

\begin{figure}[t,h]
\centerline{\psfig{figure=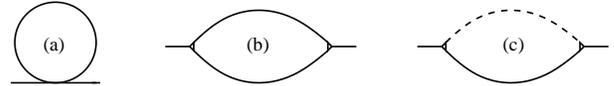,width=80mm}\vspace*{5mm}}
\caption{One-loop diagrams that renormalize $H$ and $G$.}
\label{fig:5}
\end{figure}

We now consider the one-loop renormalizations of the other vertices in
the effective action. Of particular interest are the coupling constants
$H$ and $G$ in the two-point $q$-vertex, which determine the specific
heat coefficient, and the conductivity, respectively, see
Sec.\ \ref{subsubsec:II.D.3} above. The topological structure of the
relevant diagrams is shown in Fig.\ \ref{fig:5}.
From extensive work on the nonlinear sigma model it is known that diagram
(b), combined with suitable contributions from diagram (a),
is finite for $d>2$.\cite{us_R}
A calculation of diagrams (c) and the remaining parts of (a) reveals
that their contribution to the one-loop renormalization of $G$
consists of two pieces which, at zero external frequency, are given by
\bml
\label{eqs:3.6}
\be
({\delta G})_1 = \frac{3}{8}\,G^3\,\frac{K_t}{V} 
   \sum_{\bf p} 2\pi T\sum_{l=0}^{\infty}
     {\cal D}_l^2({\bf p})\,{\cal M}_l({\bf p})\quad,
\label{eq:3.6a}
\ee
and
\bea
(\delta G)_2 &=& \frac{3}{8}\,G^3\,K_t\,
     \frac{\partial}{\partial {\bf k}^2} \biggl\vert_{{\bf k}=0}
   \frac{1}{V}\sum_{\bf p} 2\pi T\sum_{l=0}^{\infty} {\cal M}_l({\bf p})
\nonumber\\
&&\hskip 50 pt \times\left[
      {\cal D}_l({\bf p}) - {\cal D}_l({\bf p}+{\bf k})\right]\ .
\label{eq:3.6b}
\eea
\eml%
respectively.
Both of these integrals are finite in $d>2$. A simple calculation
shows that the one-loop correction to the density of states is given
by the same integral as $(\delta G)_1$.
For the $H$-renormalization,
diagram (b) is again finite, while the other two contributions yield
\bml
\label{eqs:3.7}
\be
\delta H = \frac{3}{8}\,G\,K_t\,\frac{1}{\Omega_n}\,2\pi T\sum_{l=0}^{n}
   \frac{1}{V}\sum_{\bf p}{\cal D}_l({\bf p})\,{\cal M}_l({\bf p})\quad,
\label{eq:3.7a}
\ee
where $n$ is the external frequency label.
For later reference we note that diagrams (a) and (c) in Fig.\ \ref{fig:5}
each contribute one half of this result. Individually, each of these
diagrams also contributes pieces that diverge like $1/\Omega_n$; these
contributions cancel between the two diagrams. (The same is true for
diagram (b) and the corresponding contributions from diagram (a).)
Notice that the frequency structure is slightly different than in the
case of the $G$-renormalization. This leads to a finite frequency sum
in Eq.\ (\ref{eq:3.7a}), and as a result the integral is logarithmically
infrared divergent for all dimensions $2\leq d<4$,
\be
\delta H = \frac{3}{4}\,{\bar G}\,K_t\,\ln (1/\Omega_n)\quad.
\label{eq:3.7b}
\ee
Here ${\bar G} = G\,S_d/(2\pi)^d$, with $S_d$ the surface area of the
unit $(d-1)$-sphere.
While this is consistent with the result of Ref.\ \onlinecite{us_dirty} that
the specific heat coefficient at the quantum ferromagnetic transition
is logarithmically divergent for all $2<d<4$, it is obvious from the
present formulation of the theory that this is very unlikely to be 
the exact critical
behavior as claimed in that reference. The reason is that an insertion
of the one-loop $H$ into the one-loop diagrams yields a $\ln^2\Omega$
singularity, and so on. Similarly, inserting the one-loop result for
the $q$-propagator in the diagram shown in Fig.\ \ref{fig:4} generates
a logarithmic correction to the vertex $u_2({\bf k})$.
Unless these insertions are exactly cancelled
by skeleton diagram contributions, the exact critical behavior must
therefore involve a more complicated function of $\ln\Omega$. We will
come back to this point in the next subsection, and in II.

Finally, the diagrams shown in Fig.\ \ref{fig:5} also determine the
renormalization of $K_s$. Again, diagram (b) yields a finite result,
while from the other two diagrams one expects
\be
\delta K_s = -\delta H\quad,
\label{eq:3.7c}
\ee
\eml%
with $\delta H$ the divergent part of the $H$-renormalization,
Eq.\ (\ref{eq:3.7b}). This result follows from what is known
about the nonlinear sigma model as we will discuss in
Sec.\ \ref{subsec:IV.A}.

Notice that the effects we have discussed above at one-loop level
would of course also occur if we had worked with the 
theory that one obtains if one puts the bare $a_{d-2}$ equal to
zero. However, in that theory their derivation would have required
a renormalization of the paramagnon propagator. In a diagrammatic
language, by adding the $\vert{\bf k}\vert^{d-2}$ term to our bare
action, we have made use of skeleton diagrams that contain certain
infinite resummations. We note in passing that one might worry about
higher orders in the loop expansion producing even stronger nonanalyticities
than $\vert{\bf k}\vert^{d-2}$. We will show in II that this is not
the case.

\subsection{Naive fixed points and their instability}
\label{subsec:III.B}

We now proceed to perform a power counting analysis of our effective
action, Eqs.\ (\ref{eqs:2.25}). Our goal is to understand the perturbative
results of the preceding subsection from a more general point of view,
and to determine the minimal effective action which, when solved, will
yield the {\em exact} critical behavior. For this purpose it is convenient 
to rewrite the action in a schematic form which suppresses everything
that is not necessary for power counting,
\bea
{\cal A}_{\rm eff}[M,q]&=&-\int d{\bf x}\ M\,\left[t + a_{d-2}\,
   \partial_{\bf x}^{d-2} + a_2\,\partial_{\bf x}^2\right]\,M
\nonumber\\
  && \hskip 60pt + O(\partial_{\bf x}^4\, M^2, M^4)
\nonumber\\
&&\hskip -30pt -\frac{1}{G}\int d{\bf x}\,(\partial_{\bf x} q)^2 
     + H\int d{\bf x}\,\Omega\,q^2 + K_s\,T\int d{\bf x}\, q^2
\nonumber\\
&& \hskip -0pt - \frac{1}{G_4}\int d{\bf x}\,\partial_{\bf x}^2\, q^4
   + H_4\int d{\bf x}\,\Omega\,q^4
\nonumber\\
&& \hskip 50pt + O(T q^3, \partial_{\bf x}^2\, q^6,
      \Omega\,q^6)
\nonumber\\
&& + \sqrt{T}\,c_1 \int d{\bf x}\,M\,q + \sqrt{T}\,c_2 \int d{\bf x}\,M\,q^2
\nonumber\\
   && \hskip 50pt + O(\sqrt{T}Mq^4)\quad.
\label{eq:3.8}
\eea
Here the fields are understood to be functions of position and frequency,
and only quantities that carry a scale dimension are shown. The bare values
of $G_4$ and $H_4$ are proportional to those of $G$ and 
$H$.\cite{renormalization_footnote} The term of order $Tq^3$, which arises
from the interacting part of ${\cal A}_{{\rm NL}\sigma{\rm M}}$, will not be
of importance for our purposes although its coupling constant squared has 
the same scale dimension as $1/G_4$ and $H_4$. It is therefore not shown
explicitly. The $c_1$,
$c_2$, etc. are the coupling constants of the terms contained in 
${\cal A}_{\rm c}$, Eq.\ (\ref{eq:2.25c}). 

We now assign a length $L$ a scale 
dimension $[L]=-1$. Under a renormalization group transformation that
involves a length rescaling by a factor $b$, all quantities will then
change according to $A \rightarrow b^{[A]}\,A$, with $[A]$ the scale
dimension of $A$. In particular, imaginary time $\tau$ and temperature
$T$ or frequency $\Omega$ have scale dimensions $[\tau] \equiv -z$,
and $[T] = [\Omega] \equiv z$, respectively.

\subsubsection{Hertz's fixed point}
\label{subsubsec:III.B.1}

To illustrate an important point, let us first show how one
recovers Hertz's mean-field fixed point\cite{Hertz} within the present 
formalism. Let us look for a fixed point where the coefficients $a_2$ and
$c_1$ are marginal, $[a_2] = [c_1] = 0$. This choice\cite{FP_footnote} 
is motivated by the
desire to find a fixed point with mean field-like static critical behavior,
and with dynamics given by the frequency dependence of the standard
paramagnon propagator ${\cal M}$, Eq.\ (\ref{eq:3.2c}). (Recall that the 
frequency dependence of ${\cal M}$ was produced by the vertex with coupling
constant $c_1$.) From the condition that the
action be dimensionless we then obtain the scale dimensions of the
order parameter field,
\bml
\label{eqs:3.9}
\be
[M_n({\bf x})] = (d-2)/2\quad,
\label{eq:3.9a}
\ee
and we find $t$ to be relevant with scale dimension
\be
[t] = 2\quad.
\label{eq:3.9b}
\ee
The correlations of the $q$-field we expect to describe the diffusive
dynamics of the fermions, so we choose\cite{us_fermions}
\be
[q_{nm}({\bf x})] = (d-2)/2\quad,
\label{eq:3.9c}
\ee
and $G$, $H$, and $K_s$ are all dimensionless.
The marginality of the coupling constant $c_1$ then implies
\be
z = [T] = 4\quad.
\label{eq:3.9d}
\ee

This is the fixed point proposed by Hertz,\cite{Hertz} which leads to
mean-field critical behavior. It is unstable because the coupling
$a_{d-2}$ is relevant with respect to this fixed point, as has been
pointed out in Ref.\ \onlinecite{us_dirty}. While this is obvious from
the action as formulated here, the following interesting question arises.
Suppose we had not added the term with coupling constant $a_{d-2}$ to 
our bare action. Since this term was generated by means of the
$Mq^2$-vertex with coupling constant $c_2$, see Fig.\ \ref{fig:4},
$c_2$ should be relevant with respect to Hertz's fixed point. However,
power counting with the above scale dimensions yields 
\be
[c_2] = \frac{-1}{2}\,(d+z-6)\quad,
\label{eq:3.9e}
\ee
\eml%
so with the above value $z=4$, $c_2$ seems to be irrelevant. The 
resolution of this paradox lies in the fact that there is more
than one time scale in the problem, and hence all factors of $T$
do not carry the same scale dimension $z$. This is obvious if we
consider the fermionic sector of our action; the factors of $T$
in the $q^2$-vertex carry a scale dimension $[T] = z_{\rm diff} = 2$,
which corresponds to the diffusive time scale that describes the
dynamics of the electronic soft modes and which is distinct from
the critical time scale that corresponds to $[T] = z_{\rm c} = 4$.
The scale dimensions of the factors of $\sqrt{T}$ in the coupling
part of the action are therefore not {\em a priori} clear, and they
may depend on the diagrammatic context a vertex is used in. Consider
the diagram in Fig.\ \ref{fig:4} again. In this context, the two factors 
of $\sqrt{T}$ contribute to the frequency measure of a fermionic loop,
and hence they carry the diffusive time scale. Indeed, with $z=2$
Eq.\ (\ref{eq:3.9e}) shows that $c_2$ is relevant for $2<d<4$, 
and its scale dimension is consistent with that of $a_{d-2}$.

It is worthwhile to mention that one can tell {\em a priori} 
that the scale dimension for $t$, Eq.\ (\ref{eq:3.9b}), cannot
correspond to a stable fixed point since it corresponds to a
correlation length exponent $\nu=1/[t]=1/2$, which violates the
Harris criterion inequality $\nu\geq 2/d$.\cite{Harris} The
relevance of $c_2$ provides an explicit mechanism for the
instability.

We also note that the above discussion is oversimplified in that it
pretends that $M$ has always the same scale dimension, independent
of the context the order parameter fluctuations appear in. As we
will see in the next subsection, this is not quite true. However,
since this point is not crucial for the instability of Hertz's
fixed point we have suppressed it.

\subsubsection{A marginally unstable fixed point}
\label{subsubsec:III.B.2}

Given the presence of the term with coupling constant $a_{d-2}$, an
obvious attempt to find a stable fixed point is to choose $a_{d-2}$
and $c_1$ to be marginal instead of $a_2$ and $c_1$.\cite{local_footnote_2}
A slight complication, however, lies in the fact that due to the existence
of two time scales, $a_{d-2}$ will not necessarily be marginal under all
circumstances. Namely, if the frequency in the paramagnon propagator,
Eq.\ (\ref{eq:3.2c}), is diffusive, i.e. if it scales like ${\bf k}^2$,
then $a_{d-2}$ will be irrelevant. As we will see below, this can happen
if the paramagnon propagator appears as an internal propagator
in perturbation theory, although in the critical paramagnon $a_{d-2}$ is
marginal. In general, we therefore demand only that $c_1$ be marginal,
that the scale dimension of the $q$-field be consistent with a diffusive
$\langle qq\rangle$ propagator,
\bml
\label{eqs:3.10}
\be
[q_{nm}({\bf x})] = \frac{1}{2}\,(d-2)\quad,
\label{eq:3.10a}
\ee
and that the diffusive time scale be represented by a
dynamical critical exponent
\be
z_{\rm diff} = 2\quad.
\label{eq:3.10b}
\ee
Equation (\ref{eq:3.8}) then implies
\be
[G] = [H] = [K_s] = 0\quad.
\label{eq:3.10c}
\ee
\eml%

The marginality of $c_1$ implies for the scale dimension of the order 
parameter field 
\bml
\label{eqs:3.11}
\be
[M_n({\bf x})] = 1 + (d-z)/2\quad,
\label{eq:3.11a}
\ee
with $z$ the dynamical exponent associated with the $\sqrt{T}$ prefactor
in the $c_1$-vertex. In the critical paramagnon propagator we expect
$a_{d-2}$ to be marginal, which implies $[M_n({\bf x})] = 1$, and hence
a critical time scale characterized by
\be
z_c = d\quad,
\label{eq:3.11b}
\ee
and a critical exponent $\eta$, defined by $[M_n({\bf x})] = (d-2+\eta)/2$,
\be
\eta = 4-d \quad.
\label{eq:3.11c}
\ee
\eml%
This makes $a_2$ irrelevant, while $t$ is relevant with scale dimension
\bml
\label{eqs:3.12}
\be
[t] = d-2\quad,
\label{eq:3.12a}
\ee
which leads to a correlation length exponent
\be
\nu = 1/[t] = 1/(d-2)\quad.
\label{eq:3.12b}
\ee
\eml%
Notice that in contrast to the situation at Hertz's fixed point, this
result respects the Harris criterion.\cite{Harris}
Here and in the remainder of this paper we restrict ourselves to the range
of dimensions $2<d<4$, which includes the physically interesting case
$d=3$. For the behavior in higher dimensions, see Ref.\ \onlinecite{us_dirty}.

The preceding results characterize the Gaussian fixed point that was discussed
in Ref.\ \onlinecite{us_dirty}. If all other terms in the action
were irrelevant, or marginal leading to finite renormalizations only,
then this fixed point would be stable. To check this, we need to consider
the corrections to the Gaussian action. We start with $c_2$, whose
scale dimension is
\be
[c_2] = 1 - z/2\quad,
\label{eq:3.13}
\ee
with $z$ the scale dimension of the factor of $T$ in that vertex. If
this temperature represents the critical time scale, then $c_2$ is
irrelevant. However, if it represents the diffusive times scale, then
it is marginal. This can indeed happen, as we have discussed in
Sec.\ \ref{subsubsec:III.B.1} above. The example we used, viz. the
diagram in Fig.\ \ref{fig:4}, just leads to a finite renormalization
of the coefficient $a_{d-2}$, which is part of our effective action
anyway. If this were the only effect of $c_2$, then we
could neglect it. However, this is not the case. The one-loop renormalization
of $H$ that was discussed in Sec.\ \ref{subsubsec:III.A.2} provides an
example how operators that appear irrelevant by naive power counting
can be effectively marginal due to the existence of two time scales,
lead to logarithms,  and therefore need to be kept. Consequently, $c_2$ is
not necessarily harmless even if $z=d$ in Eq.\ (\ref{eq:3.13}). 
This is an important point which we now discuss in detail.

Consider diagrams (a) and (c) in Fig.\ \ref{fig:5}. They both lead to
a correction to the two-point $q$-vertex that is of the form, at zero
external wavenumber and frequency,
\bml
\label{eqs:3.14}
\be
\delta\Gamma^{(2)} \propto \frac{1}{V}\sum_{\bf p}\ T\sum_{l=1}^{\infty}
  {\cal D}_l({\bf p})\,{\cal M}_l({\bf p})\quad.
\label{eq:3.14a}
\ee
For scaling purposes, let us cut off the momentum integral in the infrared
by $1/b$, with $b$ a RG length scale factor. Doing the integral then shows
that it is given by a constant plus a term proportional to $b^{-d}\,\ln b$.
The constants cancel between the two diagrams, and we have
\be
\delta\Gamma^{(2)} \sim b^{-d}\,\ln b\quad.
\label{eq:3.14b}
\ee
Notice that the frequency in the above integral scales like a wavenumber
to the power $d$, so $(c_2)^2$ in diagram (c) has a negative scale dimension
$-(d-2)$, and so does the quartic vertex in diagram (a). The salient
point is now as follows. For the purpose of the renormalization of $G$,
i.e. the wavenumber dependent part of $\Gamma^{(2)}$, we need to replace
$1/b$ by ${\bf k}$. We then obtain the gradient squared of the bare
vertex times a factor $\vert{\bf k}\vert^{d-2}$. The contribution is
therefore irrelevant, in agreement with the negative scale dimensions of
the vertices and the result of the explicit perturbative calculation. 
However, for the purpose of the renormalization of $H$ we need to replace
$1/b$ by an appropriate power of the frequency. This can be $\Omega^{1/2}$,
if $\Omega$ represents the diffusive frequency scale, or $\Omega^{1/d}$, if
it represents the critical one. Since the frequency in the integral scales
like $\vert{\bf p}\vert^d$, the latter applies and we have
\be
\delta\Gamma^{(2)} \propto \Omega\,\ln\Omega\quad,
\label{eq:3.14c}
\ee
\eml%
in agreement with Eq.\ (\ref{eq:3.7b}).

The point illustrated above is as follows. Due to the existence of two
different time scales, the fact that an operator has a negative scale
dimension by naive power counting, which is based on the consideration
of length scales, does not necessarily imply that it will be irrelevant.
Rather, operators with scale dimensions between zero and $-(d-2)$
may act as marginal operators with respect to frequency scaling. Notice,
however, that for this mechanism to be operative it is crucial that
the vertex being renormalized is proportional to frequency. Therefore,
the seemingly irrelevant operators become effectively marginal with respect 
to $H$, but not with respect to $G$ or any other coupling constant. In
the appendix we discuss another aspect of this phenomenon.

We conclude that the Gaussian fixed point of Ref.\ \onlinecite{us_dirty}
is not stable since there are operators that are effectively marginal
with respect to it. If these operators just led to finite renormalizations,
this would still not change the conclusions of the earlier paper. However,
as we have seen above, they lead to logarithmic corrections to power-law
scaling and hence need to be kept. The problem is less severe than in the
case of Hertz's fixed point, however, since now there are no relevant
operators. If one can show that all other terms are truly irrelevant,
then the conclusion would be that to determine the exact critical
behavior it suffices to keep the Gaussian action plus the $M\,q^2$ coupling
and all terms up to $O(q^4)$. We investigate this hypothesis next.

\subsection{Effective action for the critical behavior}
\label{subsec:III.C}

\subsubsection{The effective action}
\label{subsubsec:III.C.1}

From the discussion in the preceding subsection we infer an educated
guess for an effective action that contains only the terms needed for
a description of the critical fixed point and the associated critical
behavior. This action should contain all of the terms that are
shown explicitly in Eq.\ (\ref{eq:3.8}), except that for $2<d<4$ one
can drop the gradient squared term in the LGW part of the action.
Notice that we need to keep the terms of
$O(\partial_{\bf x}^2\,q^4,\Omega\,q^4)$ in the expansion of the 
nonlinear sigma model in
powers of $q$, as they give rise to diagram (a) in Fig.\ \ref{fig:5}.
These terms appear irrelevant by naive power counting, but contribute
to the leading frequency dependence by means of the mechanism discussed
in Sec.\ \ref{subsubsec:III.B.2} and in the appendix. By the same
argument one should keep the terms of order $O(q^3)$ and $O(q^4)$
that arise from the spin-singlet interaction. However, by themselves
these vertices give only rise to diagrams that are finite in 
$d>2$,\cite{us_R} and combined with $c_2$ or other vertices that
contain $M$ they lead to mixed $\langle bq\rangle$ 
propagators, Eq.\ (\ref{eq:3.4}),
which are less infrared divergent then the second term on the right-hand
side of Eq.\ (\ref{eq:3.3c}). These terms can therefore safely be neglected.
This leaves the spin-singlet interaction constant $K_s$ entering the effective
theory via the vertex $\Gamma^{(2)}$ only. Since $K_s\neq 0$ does not change
the diffusive structure of the noninteracting $q$ propagator, it can be
dropped there as well.
Restoring all indices, the suggested effective action for describing the
critical fixed point reads\cite{FP_action_footnote}
\bml
\label{eqs:3.15}
\bea
{\cal A}_{\rm FP}&=&-\sum_{{\bf k},n,\alpha}\sum_{i=1}^3
 {^iM}_n^{\alpha}({\bf k})\,\left[t + a_{d-2}\,\vert{\bf k}\vert^{d-2}\right]\,
   {^iM}_{-n}^{\alpha}(-{\bf k})
\nonumber\\
&&- \frac{4}{G}\sum_{\bf k}\sum_{1,2,3,4}\sum_{r,i}{^i_rq}_{12}({\bf k})\,
        {^i\Gamma}_{12,34}^{(2)}({\bf k})\,{^i_rq}_{34}(-{\bf k})
\nonumber\\
&&-\frac{1}{4G}\sum_{1,2,3,4}\ \sum_{r,s,t,u}\ \sum_{i_1,i_2,i_3,i_4}
   \frac{1}{V}
   \sum_{{\bf k}_1,{\bf k}_2,{\bf k}_3,{\bf k}_4}
\nonumber\\
&&\hskip 10pt\times {^{i_1i_2i_3i_4}_{\ \ \ rstu}\Gamma}
    ^{(4)}_{1234}({\bf k}_1,{\bf k}_2,{\bf k}_3,{\bf k}_4)\ 
    {^{i_1}_rq}_{12}({\bf k}_1)\,{^{i_2}_sq}_{32}({\bf k}_2)
\nonumber\\
&&\hskip 100 pt \times {^{i_3}_tq}_{34}({\bf k}_3)\,{^{i_4}_uq}_{14}({\bf k}_4)
\nonumber\\
&&+ c_1\,\sqrt{T}\sum_{\bf k}\sum_{12}{^i_rb}_{12}({\bf k})\,{^i_rq}_{12}
      (-{\bf k})
\nonumber\\
&&+ c_2\,\sqrt{T}\,\frac{1}{\sqrt{V}}\sum_{{\bf k},{\bf p}}\sum_{n_1,n_2,m}
  \sum_{r,s,t}\sum_{i=1}^3\sum_{j,k}\sum_{\alpha,\beta} 
     {^i_rb}_{n_1n_2}^{\alpha\alpha}({\bf k})\,
\nonumber\\
&&\hskip -25 pt \times\left[{^j_sq}_{n_2m}^{\alpha\beta}({\bf p})\,
  {^k_tq}_{n_1m}^{\alpha\beta}(-{\bf p}-{\bf k})\,
   \tr\left(\tau_r\tau_s\tau_t^{\dagger}\right)\,\tr\left(s_is_js_k^{\dagger}
                                                           \right)\right.
\nonumber\\
&&\hskip -20pt \left. - {^j_sq}_{mn_2}^{\beta\alpha}({\bf p})\,
  {^k_tq}_{mn_1}^{\beta\alpha}(-{\bf p}-{\bf k})\,
      \tr\left(\tau_r\tau_s^{\dagger}\tau_t\right)\,\tr\left(s_is_j^{\dagger}
                                             s_k\right)\right]\,,
\nonumber\\
\label{eq:3.15a}
\eea
with $\Gamma^{(2)}$ from Eqs.\ (\ref{eq:3.1c}) and (\ref{eq:3.1d}) 
with $K_s=0$, and
\bea
{^{i_1i_2i_3i_4}_{\ \ \ rstu}\Gamma}
    ^{(4)}_{1234}({\bf k}_1,{\bf k}_2,{\bf k}_3,{\bf k}_4) 
  &=&-\delta_{{\bf k}_1+{\bf k}_2+{\bf k}_3+{\bf k}_4,0}\,
\nonumber\\
&&\hskip -100pt\times
 \tr\left(\tau_r\tau_s^{\dagger}\tau_t\tau_u^{\dagger}\right)\,
   \tr\left(s_{i_1}s_{i_2}^{\dagger}s_{i_3}s_{i_4}^{\dagger}
              \right)\,
     \left({\bf k}_1\cdot{\bf k}_3 + {\bf k}_1\cdot{\bf k}_4\right.
\nonumber\\
&&\hskip -75pt \left. + {\bf k}_1\cdot{\bf k}_2 + 
   {\bf k}_2\cdot{\bf k}_4 - GH\Omega_{n_1-n_2} \right)\quad.
\label{eq:3.15b}
\eea
The bare values of the coupling constants $c_1$ and $c_2$ are related,
and given by
\be
c_1 = 16\,c_2 = 4\sqrt{\pi\,K_t}\quad.
\label{eq:3.15c}
\ee
\eml%
Notice that this action is {\em not} Gaussian, and therefore the critical
behavior is not easy to determine.

We will solve the effective model given by Eqs.\ (\ref{eqs:3.15}) 
in II,\cite{us_paper_II} where we will show that the exact critical
behavior differs from the Gaussian one by logarithmic corrections only.
In the remainder of this paper we show that the action given by 
Eqs.\ (\ref{eqs:3.15}) really is sufficient for describing the critical 
behavior in $2<d<4$.

\subsubsection{Corrections to the effective action}
\label{subsubsec:III.C.2}
 
We now show that all terms that were neglected in writing 
Eqs.\ (\ref{eqs:3.15}) are irrelevant by power counting, keeping in mind the
complications due to the two time scales that were dicussed in
Sec.\ \ref{subsec:III.B} above. In addition to the scale 
dimensions of $M$ and $q$ given
in Eqs.\ (\ref{eq:3.10a}), (\ref{eq:3.11a}), we need for this purpose the
scale dimension of the massive fields $\Delta P$ and $\Delta\Lambda$. 
The correlations of
$\Delta P$ are short ranged, and of the same nature as at the Fermi liquid
fixed point that was discussed in Ref.\ \onlinecite{us_fermions}. We thus
choose
\be
[\Delta P({\bf x})] = [\Delta\Lambda({\bf x})] = d/2\quad.
\label{eq:3.16}
\ee
Power counting now proceeds as usual. The nonlinear sigma model action
we have kept up to $O(q^4)$. Higher order corrections have
the same scale dimensions as
at the Fermi liquid fixed point in Ref.\ \onlinecite{us_fermions}. They
thus are all irrelevant with scale dimensions that are smaller than
$-(d-2)$ and are therefore harmless. The
couplings between $M$, $q$, and the massive modes given in 
Eq.\ (\ref{eq:2.28}), for even powers of $q$, have scale dimensions
\be
[d_n] = -\,\frac{n-1}{2}\,(d-2) - \frac{z}{2} \leq \frac{-1}{2}\,(d+z-2)\quad,
\label{eq:3.17}
\ee
for $n\geq 2$, and thus can safely be neglected.
For odd powers of $q$, the couplings contains an effective external
frequency, and therefore are even less irrelevant than Eq.\ (\ref{eq:3.17})
suggests. In particular we confirm 
that $d_1$, which we have dropped,\cite{MDPq_footnote}
has a scale dimension $[d_1] = -3z/2$ and is thus more irrelevant then
$d_2$. $[d_0] = (d-2-z)/2$, which becomes marginal in $d=4$ if
$z=z_{\rm diff}=2$. However, $d=4$
is a special dimension anyway, and for $d>4$ one
obtains a different fixed point since the $\vert{\bf k}\vert^{d-2}$ term
in the LGW action is no longer leading.
A remaining question is whether the formally irrelevant $d_0$ can be
promoted to marginal or relevant status by the same mechanism that is
operative for, e.g., $c_2$. The answer is negative, since the mechanism
works only for the renormalization of $H$, and in order to renormalize $H$,
$d_0$ needs to be combined with some $d_n$ with $n\geq 2$. However,
$[d_0\,d_2] = -z < -(d-2)$ for $d<4$. Therefore, all of the $d_n$ can
be safely neglected.
Similarly, all terms of higher than quadratic order in $M$ are irrelevant.
We mention, however, that in the ordered phase the term of $O(M^4)$ becomes
dangerously irrelevant and needs to be kept. This will be important in II. 

Finally, the random mass term ${\cal A}_{\rm LGW}^{(4,2)}$, 
Eq.\ (\ref{eq:2.23}),
deserves an extra discussion. The scale dimension of the coupling constant
$v_4$ in Eq.\ (\ref{eq:2.23}) is $[v_4] = d-4$, while the 
scale dimension of the combination of $d_0$ and $d_2$ in Eq.\ (\ref{eq:2.28})
which produce $v_4$ in perturbation theory (see Fig.\ \ref{fig:2}) is
$[(d_0\,d_2)^2] = -2z = -4$. $[v_4]$ is thus much 
less irrelevant than one might
expect from naive power counting. The resolution of this discrepancy is as
follows. The $v_4$-vertex shown in Fig.\ \ref{fig:2} at zero wavenumber 
has the schematic structure
\be
\int d{\bf y}\int d\omega\,\omega\, \langle q^2({\bf x})\,q^2({\bf y})\rangle
   \quad,
\label{eq:3.18}
\ee
which has a naive scale dimension of $d$ (with $z=2$). However, the integral
is a finite number, and so its actual scale dimension is zero. If we consider
the vertex function at a finite wavenumber $\vert{\bf k}\vert$ and perform
a gradient expansion, then we obtain an expansion of the form
\be
{\rm const.} + {\bf k}^2 + \vert{\bf k}\vert^d\quad.
\label{eq:3.19}
\ee
What happens here is that
power counting yields the scale dimension of the first {\em nonanalytic}
term in the gradient expansion, but misses more dominant analytic 
contributions. This is of no consequence as long as the latter just
renormalize existing terms in the action. Here, however, they produce a
{\em new} term in the action, viz. the random mass term, and therefore need
to be taken into account. The difference between the naive scale
dimension of the integral, Eq.\ (\ref{eq:3.18}), viz. $d$, and its actual
scale dimension, viz. zero, is precisely the difference between
$[v_4] = d-4$ and $[(d_0\,d_2)^2] = -4$.

We finally come back to the simplifications inherent in our starting point,
Eqs. (\ref{eqs:2.2}), which describes the paramagnetic phase as a 
disordered electron fluid while neglecting band structure and other
features of solids. The justification for these simplifications is as
follows. The disordered Fermi liquid fixed point
is characterized by relatively few parameters.\cite{us_fermions} 
This is in contrast to a clean Fermi liquid, which 
requires an infinite number of Fermi liquid parameters, or a whole
function, to completely characterize the fixed point.\cite{Shankar}
The crucial physical
distinction is that for the disordered case the slowest, and therefore
dominant, modes
are diffusive and arise only from electron number density, spin density, 
and particle-particle density
variables. In contrast, in the clean case there are an infinite number of
soft single-particle and two-particle modes. This simplification for the
disordered case carries over to the description of the ferromagnetic quantum
phase transition.

\section{Discussion}
\label{sec:IV}

As we have seen in Sec.\ \ref{subsec:III.C}, the effective action
for the critical behavior is
not Gaussian, and therefore a determination of the critical behavior is
nontrivial. It turns out that the critical behavior in all dimensions $d>2$
can nevertheless be determined exactly, and is given by the power laws
found in Ref.\ \onlinecite{us_dirty} with additional logarithmic corrections
to scaling. This solution of the effective
action will be deferred to II.\cite{us_paper_II}
Here we restrict ourselves to a discussion of some general features of
our effective theory, and of its relation to previous approaches to the
problem.

\subsection{Relation to other approaches}
\label{subsec:IV.A}

Let us briefly discuss the relation between our current approach and
previous theories. This is most easily done by starting from
Eq.\ (\ref{eq:2.10b}). By formally integrating out the fermions, i.e.
the fields $Q$ and $\wt\Lambda$, from this formulation of the action
one obtains an LGW theory or action entirely in terms of the order
parameter field. If the fermions are integrated out in tree
approximation, one recovers Hertz's theory.\cite{Hertz} If they
are integrated out formally exactly,
the vertices of the LGW functional are given in terms
of spin-density correlation functions for a `reference ensemble' or
fictitious electron system that has no bare spin-triplet interaction.
This is the theory that was analyzed in Ref.\ \onlinecite{us_dirty}.
The disadvantage of that approach is that the reference ensemble
contains soft modes, viz. the $q$, and integrating them out produces
effective vertices in the LGW theory that diverge in the limit of
small wavenumbers and frequencies. That is, one obtains a nonlocal
field theory. Furthermore, Ref.\ \onlinecite{us_dirty} performed
a power-counting analysis only, and integrating out the fermionic
degrees of freedom obscured the subtleties that arise in this
context due to the existence of the diffusive time scale in addition
to the critical one. As a result, the power counting analysis of
Ref.\ \onlinecite{us_dirty} was  
insensitive to the logarithmic corrections that we found by
means of explicit perturbative calculations in Sec.\ \ref{subsubsec:III.A.2}
and explained in Sec.\ \ref{subsubsec:III.B.2} in terms of a more 
sophisticated scaling analysis than the pure LGW theory allowed for.
Notice that in some other respects Ref.\ \onlinecite{us_dirty} was
actually more
sophisticated than the present theory. For instance, it included in
the bare action effects that require a one-loop analysis in the
present approach, e.g. the $\vert{\bf k}\vert^{d-2}$ term in the
vertex $u_2$. However, the insensitivity to logarithmic corrections
is hard to overcome within the framework of the nonlocal theory.

The relation between the present theory and Ref.\ \onlinecite{us_IFS}
is less obvious. To see it, consider Eqs.\ (\ref{eqs:2.25}) and
integrate out $M$. This yields a nonlinear sigma model with a triplet
interaction amplitude that is given by the static paramagnon propagator.
We have performed explicit calculations within this theory, and
ascertained that it yields the same results as the coupled $M$-$q$
theory discussed above, as it should. This equivalence between the
$M$-$q$ theory and the sigma model is the basis for Eq.\ (\ref{eq:3.7c}),
since within the sigma model $H+K_s$ is not singularly renormalized.\cite{us_R}
The bare nonlinear sigma model
in Ref.\ \onlinecite{us_IFS} had a point-like spin-triplet interaction
amplitude, but under renormalization the $\vert{\bf k}\vert^{d-2}$
that is characteristic of the static paramagnon is generated. It
is thus plausible that the pure nonlinear sigma model should contain
the critical fixed point for the ferromagnetic transition. However,
since the order parameter has been integrated out, the nature of
the transition is completely obscured within this approach, and a
description of the ordered phase is not possible. This is
the reason why Ref.\ \onlinecite{us_IFS} could only conclude that
the transition is of magnetic nature.\cite{IFS_footnote} We will
come back to the detailed connection between the two approaches in
II. Here we just mention that the present analysis positively
identifies the runaway flow that is encountered in the nonlinear
sigma model in the absence of any spin-flip mechanisms\cite{us_R}
as signaling the ferromagnetic transition.\cite{runaway_footnote}

We also mention that the fixed point identified in Sec.\ \ref{subsec:III.C}
above violates some of the general scaling laws obtained by 
Sachdev.\cite{Sachdev} As has been discussed in Ref.\ \onlinecite{us_dirty}
in some detail, this can be traced to the presence of dangerous irrelevant
variables, which can always invalidate general scaling arguments.\cite{Fisher}

We conclude that all of the previous approaches to the problem break down
at some level, and that the basic problem is always the same, namely a
lack of explicitness. Only a local field theory that correctly identifies
and keeps all of the soft modes allows for the explicit calculations
necessary to check more general arguments that may break down because of the
failure of hidden assumptions. Interestingly, as we will show in II,
the problem was solved technically correctly in Ref.\ \onlinecite{us_IFS},
but the missing physical interpretation rendered this result of limited value
at the time.

\subsection{Scaling issues}
\label{subsec:IV.B}

Let us finally come back to the issue of the two different time scales,
which has been crucial for a correct application of scaling ideas to
the problem. As we have seen in Sec.\ \ref{subsubsec:III.B.1}, the
implicit assumption of the existence of only one time scale, namely
the critical one, can lead to wrong conclusions if one relies strictly
on power counting arguments. Explicit loop calculations, on the other hand,
reveal the fallaciousness of the assumption by producing terms in the 
action that are inconsistent with the power counting. The point is that
the diffusive modes, whose time scale is different from the critical one,
produce long-range correlations {\em everywhere}, not just at the
critical point, as has been discussed in detail elsewhere.\cite{us_Ernst}
These long-range correlations are reflected, for instance, in the
$\vert{\bf k}\vert^{d-2}$ term in the LGW part of the action, 
Eq.\ (\ref{eq:2.25b}), which is responsible for the instability of Hertz's 
fixed point.

For the instability of the Gaussian fixed point of 
Ref.\ \onlinecite{us_dirty} a similar mechanism applies, although it 
is weaker and less obvious. As we have seen in the context of
Eqs.\ (\ref{eqs:3.14}), non-Gaussian terms that formally have a
negative scale dimension can effectively become marginal with respect
to frequency dependent coupling constants.
This ``counting accident'' can only happen for vertices that vanish
at zero frequency, and it has been analyzed from a RG point of view in
Sec.\ \ref{subsubsec:III.B.2} and in the appendix.

We will come back to these arguments in II, where we will provide both a
resummation of perturbation theory to all orders and a complete scaling
description of the exact critical behavior, including all logarithmic
corrections.

\acknowledgments
This work was
supported by the NSF under grant Nos. DMR-98-70597 and DMR-99-75259.

\appendix
\section{Consequences of two different time scales}

In this appendix we discuss two additional aspects of the crucial point
made in Sec.\ \ref{subsubsec:III.B.2}.

Let us first take a phenomenological scaling point of view. The scaling
equation for the two-point $q$-vertex function $\Gamma^{(2)}$ reads
\be
\Gamma^{(2)}({\bf k}=0,\Omega) = b^{-2}\,\gamma^{(2)}(\Omega b^2,\Omega b^d,
   c_2 b^{-(d-2)/2},\ldots)\quad.
\label{eq:A.1}
\ee
Here the ellipses denote the dependence of the scaling function 
$\gamma^{(2)}$ on all operators
that are not shown explicitly. Among these are $1/G_4$ and $H_4$, which
play the same role for scaling as $c_2^{\ 2}$ does. For simplicity we
restrict the discussion to the effect of the latter. In writing 
Eq.\ (\ref{eq:A.1}), we have allowed for a dependence on both the 
diffusive and the critical frequency scale. At zero-loop order, 
$\gamma^{(2)}$ depends only on
the former, and by putting $b=1/\sqrt\Omega$ we have
$\Gamma^{(2)}({\bf k}=0,\Omega) \propto \Omega$. At one-loop order,
it depends on the critical frequency as well, which opens the possibility
of a stronger frequency dependence proportional to $\Omega^{2/d}$. However,
the one-loop contribution has the property $\gamma^{(2)}(x,y,z) = f(yz^2)$,
which restores the linear frequency behavior of $\Gamma^{(2)}$. 
This is the same phenomenon that we have
discussed within the context of explicit perturbation theory in connection
with Eqs.\ (\ref{eqs:3.14}). Logarithmic corrections to scaling are
neglected in this simple argument.

To illustrate the same point from a RG flow equation point of view, and
at the same time see the origin of the logarithms, we absorb the frequency
or temperature factors multiplying $H$ and $c_1$ in Eq.\ (\ref{eq:3.8})
into these coupling constants by defining ${\tilde H} = H\Omega$, and
${\tilde c_1} = c_1\,\sqrt{T}$. For ${\tilde H}$, ${\tilde c_1}$, and
$c_2$ we then have the flow equations
\bml
\label{eqs:A.2}
\bea
\frac{d{\tilde H}}{d\ln b}&=&2\,{\tilde H} + {\rm const.}\times\,
    {\tilde c_1}^{\ 2}\, c_2^{\ 2}\quad,
\label{eq:A.2a}\\
\frac{d{\tilde c_1}}{d\ln b}&=&d\,{\tilde c_1}/2\quad,
\label{eq:A.2b}\\
\frac{dc_2}{d\ln b}&=&-(d-2)\,c_2/2\quad,
\label{eq:A.2c}
\eea
\eml%
plus higher loop orders. Again, $1/G_4$ and $H_4$ play a role analogous to
$c_2^{\ 2}$, and we have suppressed them for simplicity. 
The solution of this system of flow equations is
\be
{\tilde H}(b) = {\tilde H}(b=1)\,b^2 + {\rm const.}\times b^2\,\ln b\quad.
\label{eq:A.3}
\ee
In this picture, the positive scale dimensions of ${\tilde H}$ and 
${\tilde c_1}$ reflect the fact that frequency or temperature is a relevant 
variable. The critical frequency is more relevant than the diffusive one, but 
this difference is made up for by the fact that the critical frequency is
always multiplied by $c_2^{\ 2}$. In this way the formally irrelevant $c_2$
effectively acquires a marginal status. The logarithm, at one-loop order,
reflects a resonance between the scale dimensions of ${\tilde c_1}$ and $c_2$,
and represents one of the possibilities in Wegner's classification of
logarithmic corrections to scaling,\cite{Wegner_DG} as was already pointed out
in Ref.\ \onlinecite{us_dirty}. At higher loop order, however, additional
logarithmic terms appear as we will show in II.

\end{document}